\documentclass[ALICE,manyauthors]{cernphprep}
\usepackage[comma,square,numbers,sort&compress]{natbib}
\usepackage{hyperref}
\usepackage{lineno}
\usepackage{xspace}
\usepackage{amssymb}
\usepackage{amsmath,bm}
\usepackage{newtxmath}
\usepackage{mathrsfs}
\usepackage{epstopdf}
\usepackage{orcidlink}
\newcommand{\pp}           {\ensuremath{\mathrm{pp}}\xspace}

\newcommand{\PbPb}         {\mbox{Pb--Pb}\xspace}

\newcommand{\pPb}          {\mbox{p--Pb}\xspace}

\newcommand{\pt}           {\ensuremath{p_{\rm T}}\xspace}
\newcommand{\meanpt}       {\ensuremath{\left\langle p_{\mathrm{T}}\right\rangle}\xspace}

\newcommand{\dEdx}         {\ensuremath{\textrm{d}E/\textrm{d}x}\xspace}
\newcommand{\RpPb}         {\ensuremath{R_{\rm pPb}}\xspace}

\newcommand{\nineH}        {$\sqrt{s}~=~0.9$~Te\kern-.1emV\xspace}
\newcommand{\seven}        {$\sqrt{s}~=~7$~Te\kern-.1emV\xspace}
\newcommand{\twoH}         {$\sqrt{s}~=~0.2$~Te\kern-.1emV\xspace}
\newcommand{\twosevensix}  {$\sqrt{s}~=~2.76$~Te\kern-.1emV\xspace}
\newcommand{\five}         {$\sqrt{s}~=~5.02$~Te\kern-.1emV\xspace}
\newcommand{\twosevensixnn}{$\sqrt{s_{\mathrm{NN}}}~=~2.76$~Te\kern-.1emV\xspace}
\newcommand{\fivenn}       {$\sqrt{s_{\mathrm{NN}}}~=~5.02$~Te\kern-.1emV\xspace}

\newcommand{\GeV}          {\text{Ge\kern-.1emV}\xspace}
\newcommand{\MeV}          {\text{Me\kern-.1emV}\xspace}
\newcommand {\tev}      {\text{Te\kern-.1emV}\xspace}
\newcommand{\TeV} {\tev}

\newcommand{\MeVc}         {\ensuremath{\MeV/c}\xspace}
\newcommand{\mevc}{\MeVc\xspace}

\newcommand{\GeVc}         {\ensuremath{\GeV/c}\xspace}

\newcommand{\GeVmass}      {\ensuremath{\GeV/c^2}\xspace}
\newcommand{\MeVmass}      {\ensuremath{\MeV/c^2}\xspace}

\newcommand{\MeVcc}{\MeVmass}

\newcommand{\GeVcc}{\GeVmass}

\newcommand{\Lint}         {\ensuremath{\mathcal{L}_\mathrm{int}}\xspace}

\newcommand{\ITS}          {\rm{ITS}\xspace}

\newcommand{\TPC}          {\rm{TPC}\xspace}

\newcommand{\ee}           {\ensuremath{\mathrm{e^{+}e^{-}}}\xspace}

\newcommand{\kzero}        {\ensuremath{{\rm K}^{0}_{\rm{S}}}\xspace}

\newcommand {\pT}        {\pt}
\newcommand {\meanpT}    {\ensuremath{\langle p_{\mathrm{T}} \kern-0.1em\rangle}\xspace}
\newcommand {\mean}[1]   {\ensuremath{\langle #1 \kern-0.1em\rangle}\xspace}

\newcommand {\abs}[1]    {\ensuremath{\left | #1 \right |}}

\newcommand {\abspseudorap} {\mbox{$\left | \eta \right | $}}

\newcommand {\Raa}       {\ensuremath{R_\mathrm{AA}}\xspace}
\newcommand{\RAA}{\Raa}

\newcommand {\ep}        {\mbox{$\mathrm {e^-p}$}\xspace}

\newcommand {\MeanNpart} {\mbox{\ensuremath{< \kern-0.15em N_{part} \kern-0.15em >}}}

\newcommand{\nsigmaTPC} {\ensuremath{\mathrm{n_\sigma^{TPC}}}\xspace}
\newcommand{\nsigmaTOF} {\ensuremath{\mathrm{n_\sigma^{TOF}}}\xspace}

\newcommand {\mmom}     {\mbox{\rm MeV$\kern-0.15em /\kern-0.12em c$}}
\newcommand {\gmom}     {\mbox{\rm GeV$\kern-0.15em /\kern-0.12em c$}}
\newcommand {\mmass}    {\mbox{\rm MeV$\kern-0.15em /\kern-0.12em c^2$}}

\newcommand {\cm}       {\mbox{${\rm cm}$}}

\newcommand {\dg}       {\mbox{$\kern+0.1em ^\circ$}}

\newcommand{\nbinv}{\ensuremath{\rm nb^{-1}}}
\newcommand {\ubinv}{\ensuremath{\mu\rm b^{-1}}}
\newcommand{\mubinv}{\ubinv}

\newcommand{\mub}{\ensuremath{\mu\rm b}\xspace}
\newcommand{\mb}{\ensuremath{\rm mb}\xspace}

\newcommand{\ct}{\ensuremath{ct}\xspace}

\newcommand{\Lb}{\ensuremath{\rm {\Lambda_b^{0}}}\xspace}

\newcommand{\lambdac}     {\ensuremath{\mathrm {\Lambda_{c}^{+}}}\xspace}

\newcommand{\xiczp}        {\ensuremath{\mathrm {\Xi_{c}^{0,+}}}\xspace}
\newcommand{\XicD} {\ensuremath{\xiczp/\Dz}\xspace}

\newcommand{\rmLambdas}         {\ensuremath{\mathrm {\Lambda \kern-0.2em + \kern-0.2em \overline{\Lambda}}}\xspace}

\newcommand{\Kzs}               {\ensuremath{\mathrm {K^0_S}}\xspace}

\newcommand{\Dzero}{\ensuremath{\mathrm {D^0}}\xspace}
\newcommand{\Dz}{\Dzero}

\newcommand{\Dplus}{\ensuremath{\rm D^+}\xspace}

\newcommand{\Dsubs}{\ensuremath{\rm D_{s}^+}\xspace}
\newcommand{\Ds}{\Dsubs}

\newcommand{\Lcminus}{\ensuremath{\rm {\overline{\Lambda}{}_c^-}}\xspace}
\newcommand{\Lcplus}{\lambdac}
\newcommand{\Lc}         {\Lcplus}
\newcommand{\LcD} {\ensuremath{\lambdac/\Dzero}\xspace}

\newcommand{\LctopKzS}{\ensuremath{\rm \Lambda_{c}^{+}\to p K^{0}_{S}}\xspace}

\newcommand{\LctopKs}{\LctopKzS}

\newcommand{\sqrtsseven}{\ensuremath{\sqrt{s} = 7~\TeV}\xspace}
\newcommand{\sqrtsfive}{\ensuremath{\sqrt{s} = 5.02~\TeV}\xspace}
\newcommand{\sqrtsthirt}{\ensuremath{\sqrt{s} = 13~\TeV}\xspace}

\newcommand{\sqrtsNNfive}{\ensuremath{\sqrt{s_\mathrm{NN}} = 5.02~\TeV}\xspace}

\newcommand{\figref}[1]{Fig.~\ref{#1}}
\newcommand{\Figref}[1]{Figure~\ref{#1}}

\newcommand{\tabref}[1]{Table~\ref{#1}}

\newcommand{\Eqref}[1]{Eq.~\eqref{#1}}

\newcommand{\secref}[1]{Section~\ref{#1}}
\newcommand{\Secref}[1]{Section~\ref{#1}}

\newcommand{\Omegac}{\ensuremath{\Omega_{\rm c}^{0}}\xspace}
\newcommand{\Sigmac}{\ensuremath{\Sigma_{\rm c}^{0,++}}\xspace}

\newcommand{\lowptbin}{\ensuremath{0<\pt<1}~\GeVc}

\begin{document}

%%%%%%%%%%%%%%%  Title page %%%%%%%%%%%%%%%%%%%%%%%%
\begin{titlepage}
% the dates below correspond to CERN approval
% please don't touch: EB chairs will take care
\PHyear{2022}       % required, will be obtained from CERN
\PHnumber{261}      % required, will be obtained from CERN
\PHdate{18 November}  % required, will be obtained from CERN
%%%%%%%%%%%%%%%%%%%%%%%%%%%%%%%%%%%%%%%%%%%%%%%%%%%%

%%% Put your own title + short title here:
\title{First measurement of $\mathbf{\Lambda}_\mathbf{c}^\mathbf{+}$ production down to $p_\mathbf{T} = \mathbf{0}$ in pp and \pPb collisions at $\sqrt{s_\mathbf{NN}}= \mathbf{5.02~TeV}$}
\ShortTitle{\Lc production down to $\pt=0$ in pp and \pPb collisions}   % appears on left page headers

%%% Do not change the next lines
\Collaboration{ALICE Collaboration\thanks{See Appendix~\ref{app:collab} for the list of collaboration members}}
\ShortAuthor{ALICE Collaboration} % appears on right page headers, do not change

\begin{abstract}
The production of prompt \Lc baryons has been measured at midrapidity in the transverse momentum interval \lowptbin for the first time, in pp and \pPb collisions at a centre-of-mass energy per nucleon--nucleon collision \sqrtsNNfive. The measurement was performed in the decay channel \LctopKs by applying new decay reconstruction techniques using a Kalman-Filter vertexing algorithm and adopting a machine-learning approach for the candidate selection. The \pt-integrated \Lc production cross sections in both collision systems were determined and used along with the measured yields in \PbPb collisions to compute the \pt-integrated nuclear modification factors \RpPb and \Raa of \Lc baryons, which are compared to model calculations that consider nuclear modification of the parton distribution functions. The \LcD baryon-to-meson yield ratio is reported for pp and \pPb collisions. Comparisons with models that include modified hadronisation processes are presented, and the implications of the results on the understanding of charm hadronisation in hadronic collisions are discussed. A significant ($3.7\sigma$) modification of the mean transverse momentum of \Lc baryons is seen in \pPb collisions with respect to \pp collisions, while the \pt-integrated \LcD yield ratio was found to be consistent between the two collision systems within the uncertainties.

\end{abstract}
\end{titlepage}

\setcounter{page}{2} %please do not remove this line

%%%%%%%%%%%%%%%%%%%%%%%%%%%%%%%%
% begin main text
%%%%%%%%%%%%%%%%%%%%%%%%%%%%%%%%

\section{Introduction} 

Measurements of heavy-flavour hadron production in hadronic collisions provide crucial tests for calculations based on quantum chromodynamics (QCD).  Typically, calculations of \pt-differential heavy-flavour hadron production cross sections in hadronic collisions are factorised into three separate components: the parton distributions functions (PDFs), which describe the Bjorken-$x$ distributions of quarks and gluons within the incoming hadrons; the hard-scattering cross section for the partons to produce a charm or beauty quark; and the fragmentation functions, which characterise the hadronisation of a quark to a given hadron species~\cite{Collins:1989gx}. As charm and beauty quarks have masses much larger than the $\Lambda_\mathrm{QCD}$ energy scale, the parton--parton hard-scattering cross sections can be calculated perturbatively~\cite{Cacciari:1998it}. In contrast, the fragmentation functions cannot be calculated with perturbative QCD (pQCD) methods, and so must be determined from measurements in \ee collisions. They are then applied in cross section calculations under the assumption that the relevant hadronisation processes are ``universal'', i.e.~independent of the collision system. Hadron-to-hadron production ratios within the charm sector, such as $\Dsubs/\Dzero$ and \LcD, are therefore especially effective for probing hadronisation effects, since in theoretical calculations the PDFs and partonic interaction cross sections are common to all charm-hadron species and their effects almost fully cancel in the yield ratios.

Previous measurements of charm-meson production cross sections in pp collisions at the LHC~\cite{ALICE:2021mgk, ALICE:2017olh, LHCb:2016ikn, CMS:2021lab} show that the $\Dplus/\Dz$ and $\Ds/\Dz$ ratios are independent of the transverse momentum (\pt) within uncertainties, and are consistent with results from \ee and \ep collisions~\cite{Gladilin:2014tba}. The ratios are also described well by the PYTHIA 8 event generator using the Monash tune~\cite{sjostrand2008brief,Skands:2014pea}, which adopts hadronisation fractions based on fragmentation functions from \ee collisions. However, the charm baryon-to-meson ratios \LcD, \XicD, $\Omegac/\Dz$, and $\Sigmac/\Dz$ measured at midrapidity at the LHC~\cite{ALICE:2017thy, ALICE:2020wla,  ALICE:2020wfu,CMS:2019uws,ALICE:2021psx, ALICE:2021bli, ALICE:2021rzj, ALICE:OmegaC} show significant deviations from the values measured in \ee collisions, and the Monash tune of PYTHIA significantly underpredicts the production rates of charm baryons. Further hadronisation effects apart from pure in-vacuum fragmentation must therefore be considered in order for models to better describe the \Lc measurements. These effects include colour reconnection beyond the leading-colour approximation in PYTHIA 8~\cite{Christiansen:2015yqa}, quark coalescence effects such as those applied in the Catania model~\cite{Minissale:2020bif} and in the quark (re)combination model (QCM)~\cite{Song:2018tpv}, or variations of the statistical hadronisation model (SHM) including feed-down to the ground-state baryon species from the decays of yet-unmeasured resonant states predicted by the Relativistic Quark Model (RQM)~\cite{He:2019tik}. However, for the heavier charm-strange baryon states \xiczp and \Omegac~\cite{ALICE:2021bli,ALICE:OmegaC}, only the Catania model is able to adequately describe the data. Measurements of beauty-baryon production in pp collisions by the CMS and LHCb Collaborations~\cite{CMS:2012wje,LHCb:2019fns,LHCb:2015qvk} also indicate  similar differences in hadronisation mechanisms in the beauty sector between hadronic and leptonic collision systems.

Differences between leptonic and hadronic collision systems are further highlighted by the measured fragmentation fractions of ground-state single-charm hadrons, as reported at midrapidity for pp collisions at centre-of-mass energy \sqrtsfive in Ref.~\cite{ALICE:2021dhb}, where a significant enhancement of \Lc and \xiczp is seen with respect to \ee and \ep collisions, along with a corresponding depletion of the relative fraction of D mesons. However, the determination of these fragmentation fractions is dependent on model assumptions, as the evaluation of the \pT-integrated production cross sections of \Lc and \xiczp baryons required an extrapolation in order to cover regions of phase space that were not possible to study experimentally. This is especially relevant in the low-\pt region, where a significant fraction of the overall production of charm hadrons occurs and the uncertainties on the factorisation and renormalisation scales of pQCD calculations used for the extrapolation become large. Measuring down to low \pT is highly challenging, due to the smaller displacement of the decay vertex from the interaction vertex, limiting the effectiveness of topological selections due to the finite detector resolution. This necessitates the use of alternative reconstruction and selection techniques to extract a significant signal from the combinatorial background.

Charm hadrons are also studied in \pPb collisions at the LHC in order to examine possible modifications of their production due to the presence of a cold nuclear environment. The nuclear modification factor, \RpPb, of D mesons measured by ALICE in \pPb collisions at centre-of-mass energy per nucleon--nucleon collision \sqrtsNNfive is consistent with unity for $0<\pt<36~\GeVc$~\cite{ALICE:2019fhe}, suggesting that the cold nuclear matter effects that influence charm-hadron production at midrapidity are moderate. However, measurements of \Lc baryons in \pPb collisions~\cite{ALICE:2020wla} indicate a \pt-dependent modification with respect to D mesons, with an \RpPb lower than unity for $1<\pt<2~\GeVc$ and systematically above unity for $\pt>2~\GeVc$. This result is consistent with an increase in the mean \pT of charm baryons in  \pPb collisions with respect to pp collisions. Similar effects have been observed in differential studies of \Lc and \Dzero production as a function of charged-particle multiplicity in pp collisions at \sqrtsthirt by ALICE~\cite{ALICE:2021npz}, where the \pt dependence of the \LcD ratio was significantly modified in high-multiplicity collisions with respect to low-multiplicity collisions without any significant effect on the \pT-integrated \LcD ratio. This can be extended by studying highly peripheral \PbPb collisions, where the multiplicity densities of charged particles coincide with the highest multiplicity classes in pp collisions at \sqrtsthirt. The \LcD ratios measured by the LHCb Collaboration in peripheral \PbPb collisions at forward rapidity~\cite{LHCb:2022ddg} exhibit a significant \pt dependence, albeit with systematically lower values than those measured in the same \pT region at midrapidity. However, when these are calculated after integrating in the visible \pT region, they do not have any significant dependence on the number of nucleons participating in the collision, $\left\langle N_\mathrm{part}\right\rangle$, reaffirming the independence of the baryon-to-meson ratio on the multiplicity. A modification of the \pT shape as a function of multiplicity has also been observed in the strangeness sector by the ALICE and CMS Collaborations~\cite{ALICE:2013wgn, CMS:2019isl} and is consistent with the effect of radial flow in hydrodynamic models such as EPOS LHC~\cite{PhysRevC.92.034906}. In this picture, particles of larger mass are boosted to higher transverse momenta due to the presence of a common velocity field~\cite{PhysRevC.48.2462}. Furthermore, baryon production may be enhanced as a result of hadronisation by quark recombination~\cite{PhysRevLett.90.202303}. This can be further examined by extending the measurement of \Lc-baryon production down to $\pt=0$ in both collision systems and determining the mean transverse momentum. In addition, comparisons between \pPb and \PbPb collisions make it possible to disentangle initial- and final-state nuclear effects on charm-baryon production in heavy-ion collisions. The effect of nuclear shadowing~\cite{Armesto:2006ph}, which arises due to a modification of the nuclear PDFs, can lead to a reduction in the charm-hadron yields at low \pt due to a reduction of parton densities at low Bjorken-$x$. The nuclear modification factor \Raa of \Lc baryons at midrapidity in central \PbPb collisions at \sqrtsNNfive has a value systematically lower than unity for $\pt<4~\GeVc$, where nuclear shadowing is expected to play a relevant role, and $\pt>6~\GeVc$~\cite{ALICE:2021bib}, as expected from parton energy loss in the quark--gluon plasma created in the collision, while for $4<\pt<6~\GeVc$ it is consistent with unity. Measurements by the CMS Collaboration in the region $10<\pt<20$~\GeVc~\cite{CMS:2019uws} confirm this suppression at high \pT, with an indication of increased suppression for central (0--30\%) compared to peripheral (30--100\%) collisions. Studying the \pt-integrated nuclear modification factors allows us to determine whether the modification of the production yields observed in specific \pt regions is due to a reduction of the overall \Lc yield, or a modification of the momentum spectra in different collision systems.

In this article, new measurements of \Lc-baryon production in the \pt region \lowptbin in pp and \pPb collisions at \sqrtsNNfive are reported. With respect to the previously published \Lc production cross sections~\cite{ALICE:2020wfu,ALICE:2020wla}, the measurements in both systems are extended down to $\pt=0$ thanks to new decay reconstruction techniques, which employ a Kalman-Filter (KF) vertexing algorithm~\cite{kfparticle} coupled with machine-learning-based  selections~\cite{chen2016xgboost}. The \pt-integrated \Lc production cross sections and \LcD ratios reported in Ref.~\cite{ALICE:2020wla} are updated using these results, and the \pt-integrated nuclear modification factor \RpPb is calculated. The new values are obtained without requiring a model-dependent extrapolation in the \lowptbin interval. The measurement of the full momentum spectrum also enables the calculation of the mean \pt of \Lc baryons in pp and \pPb collisions. The integrated production cross section in pp collisions is used along with the measured \Lc yields in \PbPb collisions~\cite{ALICE:2021bib} in order to derive the \pt-integrated nuclear modification factor \Raa. The paper is organised as follows. \secref{sec:datasample} describes the ALICE apparatus and the analysed data samples. \Secref{sec:methods} details the analysis methods that were used. Sections~\ref{sec:corrections} and~\ref{sec:systematics} outline the corrections that are applied to calculate the \Lc production cross sections, and the sources of systematic uncertainty. The results are presented in~\secref{sec:results} and compared with model calculations. Finally, a brief summary is given in~\secref{sec:summary}.

\section{Experimental setup and data samples} \label{sec:datasample}

The ALICE detector system and its performance are described in detail in Refs.~\cite{ALICE:2008ngc,ALICE:2014sbx}. The reconstruction of charm baryons from their hadronic decay products at midrapidity primarily relies on the Inner Tracking System (ITS)~\cite{ALICE:2010tia}, the Time Projection Chamber (TPC)~\cite{Alme_2010}, and the Time-Of-Flight detector (TOF)~\cite{Akindinov:2013tea} for tracking, primary and decay vertex reconstruction, and charged-particle identification (PID). These detectors are located inside a solenoidal magnet of field strength 0.5~T. In addition, the V0 scintillator arrays~\cite{ALICE:2013axi} are used for triggering collision events and for determining the luminosity when used in conjunction with the T0 detector~\cite{ALICE:2016ovj}, and the Zero-Degree Calorimeter (ZDC) is employed for offline event rejection in \pPb collisions~\cite{ALICE:2014sbx}.

The analysis was performed at midrapidity on data from pp and \pPb collisions at \sqrtsNNfive collected with a minimum-bias (MB) trigger during Run 2 of the LHC. For pp collisions, the results are quoted for $|y|< 0.5$, whereas for \pPb collisions the rapidity in the nucleon--nucleon centre-of-mass system $(y_\mathrm{cms})$ is shifted due to the asymmetry of the colliding beams, corresponding to a rapidity range of $-0.04<y_\mathrm{cms}<0.96$. 

The MB trigger requires a pair of coincident signals in the two V0 scintillator arrays.  Further offline selections were applied to suppress the background originating from beam--gas collisions and other machine-related background sources~\cite{ALICE:2020swj}. In order to maintain uniform ITS acceptance in pseudorapidity, only events with a reconstructed vertex position within 10 cm along the beam axis from the nominal interaction point were analysed. The primary vertex position was identified using tracks reconstructed in the \TPC and \ITS detectors. Events with multiple interaction vertices due to pileup from several collisions were removed using an algorithm based on tracks reconstructed with the TPC and ITS detectors~\cite{ALICE:2014sbx}. Using these selection criteria, the sample of pp collisions comprised approximately one billion events, corresponding to an integrated luminosity of $\Lint=19.5\pm0.4~\nbinv$~\cite{ALICE-PUBLIC-2018-014}, while in \pPb collisions approximately 600 million events were selected, corresponding to $\Lint=287\pm11~\mubinv$~\cite{ALICE:2014gvw}. 

\section{Analysis methods} \label{sec:methods}

In this analysis, \Lc baryons were reconstructed via the decay channel \LctopKzS and respective charge conjugates, with branching ratio BR $= (1.59 \pm 0.08)\%$, followed by the subsequent decay $\kzero \rightarrow \pi^{+} \pi^{-}$, BR $= (69.2 \pm 0.05)\%$~\cite{pdg2022}. The contributions from both \Lc and \Lcminus were taken into account in the measurements; for brevity, both are referred to collectively as ``\Lc'' in this article.
Charged-particle tracks and particle-decay vertices were reconstructed in the central barrel using the ITS and the TPC. The particle trajectories in the vicinity of the primary vertex, and the decay vertices, were reconstructed with the KFParticle package~\cite{kfparticle}, which allows a direct estimate of their parameters and the associated uncertainties. The \kzero candidate was reconstructed by pairing opposite-sign charged tracks forming a neutral decay vertex displaced from the primary vertex. This candidate was then paired with a proton-candidate track,  originating from the primary vertex, to form a \Lc candidate. 

To ensure good quality of the tracks used to reconstruct the \Lc candidates, further selection criteria were applied in addition to the event selections mentioned above. In order to maintain a uniform detector acceptance, the tracks of the charged particles involved in the decay chain were required to be within the pseudorapidity interval
$\abspseudorap < 0.8$. The number of clusters in the \TPC used for the energy loss determination was required to be larger than 50, to enhance the precision of the mean specific energy loss (\dEdx). Furthermore, for the track reconstruction, the minimum required number of crossed rows in the \TPC was $70$ out of a possible $159$. Primary proton candidates were required to have a minimum of four (out of a maximum of six) hits in the ITS.

Several selection criteria on the PID and decay topology were applied to initially filter \Lc signal candidates. The PID selections were based on the difference between the measured and expected detector signals for a given particle species hypothesis, in units of the detector resolution ($\mathrm{n}_\sigma^\mathrm{det}$). For the pion-candidate tracks from the \Kzs decay and the proton-candidate track, a selection on the measured \dEdx in the TPC of \abs{\nsigmaTPC} $< 3$  from the respective particle hypothesis was applied. If a measurement in the TOF detector was available, a further TOF PID selection of \abs{\nsigmaTOF} $< 3$ ($5$) was applied on the particle flight time in \pPb (pp) collisions. The transverse momentum of the proton was required to be larger than $150$~$\mevc$. The deviation of the measured invariant mass from the world-average value~\cite{pdg2022} was required to be within $20$~\MeVcc for the \kzero. The \Lc candidates were also required to have a  $\chi^2_{\rm{topo}}/\rm{NDF}< 50$, where NDF is the number of degrees of freedom of the topological fit. The $\chi^2_{\rm topo}/\rm NDF$ characterises whether the momentum vector of the \Lc candidate points back to the reconstructed primary vertex, and is calculated by the KFParticle algorithm~\cite{kfparticle}.  A requirement on the distance between the primary and secondary vertices ($l$)  normalised by its uncertainty ($\Delta l$) of $l/\Delta l < 30$ was imposed on the \Lc candidate to filter out decay vertices from longer-living particles.  Finally, the estimated proper time \ct of the \kzero decay and its decay length in the transverse plane were required to be smaller than $50$~\cm.

After applying the selections described above, the separation between signal and background was optimised using a boosted decision tree (BDT) algorithm. The BDT implementation provided by the XGBoost library was used~\cite{chen2016xgboost,hipe4ml}. With the machine learning approach, multiple selection criteria are combined into a single response variable representing the probability of a candidate being a true \Lc baryon. After the application of a trained BDT model to the full data sample, a selection in the BDT response was applied to reduce the large combinatorial background.

Separate BDT models were trained for each collision system with a sample of signal and background candidates in the interval \lowptbin. The signal candidates were obtained from simulated events using the PYTHIA 8.243~\cite{sjostrand2008brief} Monte Carlo (MC) generator with the Monash tune~\cite{Skands:2014pea}. The transport of simulated particles within the detector was performed with the GEANT3 package~\cite{Brun:1082634}, and included a detailed description of the LHC beam conditions and detector geometry and alignment, as well as the time evolution of the detector configurations during the data taking. For \pPb collisions, an underlying \pPb event generated with the HIJING 1.36 generator~\cite{wang1991hijing} was added on top of the PYTHIA 8 event to simulate events with more than one nucleon--nucleon collision. Each PYTHIA 8 event was required to contain a charm--anticharm quark pair with at least one of them hadronising into a \Lc baryon. Its decay channel was then selected to be the hadronic decay into a proton and a \kzero. 
Only prompt \Lc signal candidates, namely those produced directly in the hadronisation of a charm quark or in the strong decay of a directly produced excited charm-hadron state, were selected for the training. Those that were produced in the decay of a particle containing a beauty quark (feed-down) were not used since they have a different decay vertex topology. The background sample was selected from a fraction of real data using the same filtering selections described above, with the additional requirement that the invariant mass of the \Lc candidate was within the intervals $1.98 < M < 2.23$~\GeVcc or $2.34 < M < 2.58$~\GeVcc to ensure that the signal region was excluded. 

\begin{figure}[tb]
    \begin{center}
    \includegraphics[width = 0.49\textwidth]{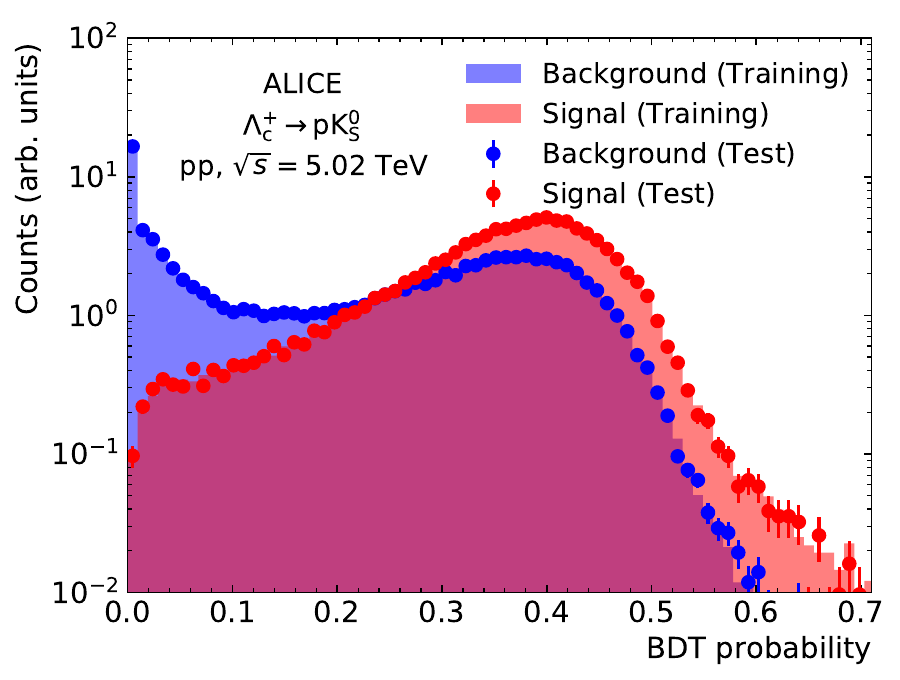}
    \includegraphics[width = 0.49\textwidth]{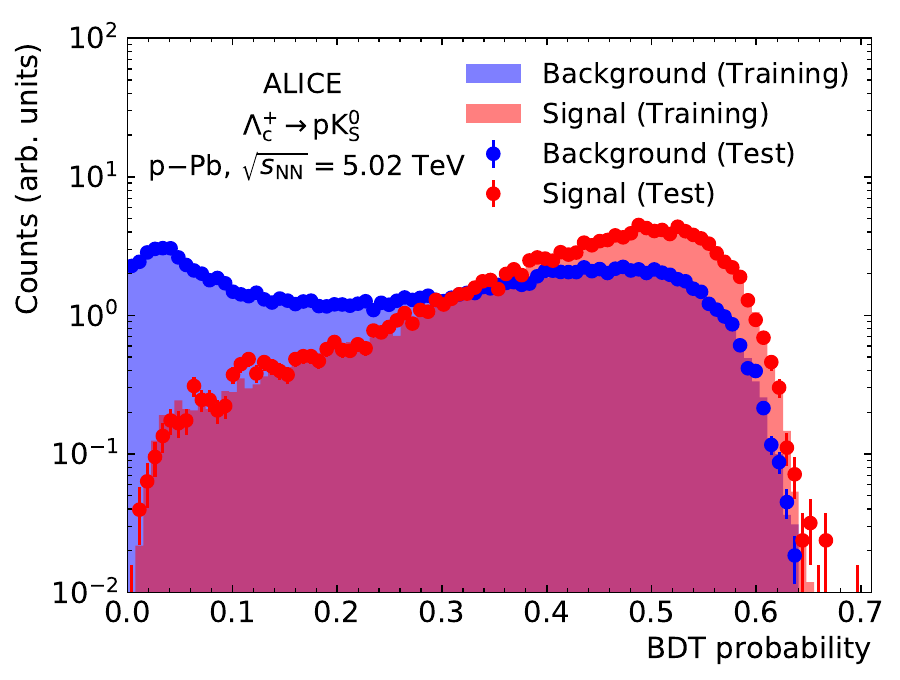}
    \end{center}
    \caption{Distributions of the BDT output probabilities for \LctopKs signal (red) and background (blue) candidates for \lowptbin. The left plot shows the model output for pp collisions, and the right plot for \pPb collisions. The shaded regions represent the output of the training sample, and the markers are the results after applying the model on the test sample.} 
    \label{fig:MLModel}
\end{figure}

The training variables related to the proton decay track were the \nsigmaTPC and the track impact parameter with respect to the primary vertex. The training variables describing the topology of the \kzero were i) the \ct, ii) the decay length in the transverse plane, and iii) the $l/\Delta l$, as defined above. The training variables related to the \Lc itself were i) the $\chi^2_{\mathrm{topo}}/\mathrm{NDF}$, ii) the $l/\Delta l$, and iii)  the pointing angle, which is defined as the angle between the momentum vector of a particle and the line connecting its production and decay vertices. \Figref{fig:MLModel} shows the BDT output probability distribution from the trained model for pp and \pPb collisions in \lowptbin, testing the hypothesis that the candidate belongs to the signal class. The normalised distributions are shown separately for the signal (red) and background (blue) classes, for the training sample (displayed as shaded bars) and the test sample (circles), which is a subset of the input data that was not used for training. The training and test distributions do not deviate significantly, demonstrating that the model is not over-trained. This was further verified using the area under the receiver operating characteristic curves~\cite{hastie_09_elements-of.statistical-learning} from the trained models, where for both collision systems a compatible value was found between the training and testing samples. In addition, while the models for the two collision systems peak at different probability values, the overall shape of the BDT output behaves similarly for pp and \pPb collisions. The proton PID variable and the \Lc $\chi^2_{\mathrm{topo}}/\mathrm{NDF}$ were found to have the highest importance ranking in the model, estimated using the SHAP package~\cite{lundberg2020local}, in both collision systems. In addition, the \ct of the \Kzs contributed significantly to the signal and background separation. Despite the limited separation of the two classes, the selection on the BDT output strongly reduces the background contribution while maintaining a high signal efficiency. The BDT probability threshold for a candidate to be selected was optimised to maximise the expected statistical significance. This was calculated using i) an estimated value for the signal in the \lowptbin region based on a L\'{e}vy-Tsallis fit to the \pt-differential \Lc production cross sections at higher \pt~\cite{ALICE:2020wla,ALICE:2020wfu}, multiplied by the reconstruction and selection efficiencies for each BDT selection threshold, and ii) an estimate of the background within the signal region obtained by interpolating a fit to the invariant mass sidebands using a fraction of the data. The resulting BDT output thresholds were 0.20 for pp collisions, and 0.37 for \pPb collisions.

\begin{figure}[tb]
    \begin{center}
    \includegraphics[width = 0.96\textwidth]{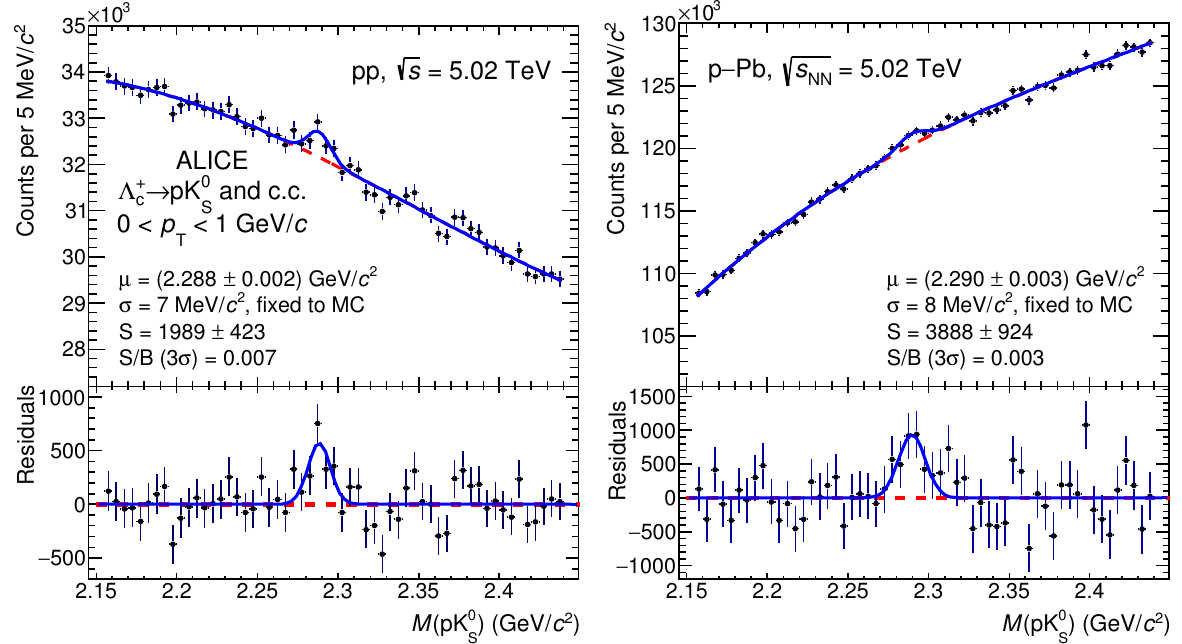}
    \end{center}
    \caption{Invariant mass distributions of \LctopKs candidates in \lowptbin, in pp (left) and \pPb (right) collisions at \sqrtsNNfive. The red dashed curves represent the background fits, and the blue curves the total fits. The lower panels show the distributions after subtracting the background estimated with the fit.}
    \label{fig:InvMassFit}
\end{figure}

After applying the BDT selections, the raw \Lc yields in the \pt interval \lowptbin were obtained by fitting the invariant-mass distributions of the candidates as shown in \figref{fig:InvMassFit}. The left (right) panel shows the invariant-mass distribution for pp (\pPb) collisions along with the fit functions. The signal peak was modelled with a Gaussian function and the background was described with a third-order polynomial. The width of the Gaussian distribution was fixed to the value obtained from MC simulations in order to improve the stability of the fit, while the mean was left as a free parameter. To better visualise the line shape of the signal, the invariant mass distributions after subtracting the background fit functions are shown in the lower panels of~\figref{fig:InvMassFit}. The statistical significance of the extracted signal has a value of $3.8$ ($3.5$) for pp (\pPb) collisions.

\section{Corrections} \label{sec:corrections}

The \pt-differential production cross section of prompt \Lc baryons per unit rapidity in the interval $|y|<0.5$ for pp collisions and $-0.96<y_\mathrm{cms}<0.04$ for \pPb collisions was calculated from the raw yields as

\begin{equation}
\frac{\text{d}^2\sigma}{\text{d}p_\mathrm{T}\text{d}y} = \frac{1}{2} \frac{f_{\mathrm{prompt}}(p_{\mathrm{T}}) \times N_\mathrm{raw}^{\Lambda^{+}_{\rm c} + \Lambda^{-}_{\rm c}}(p_\mathrm{T})}{\Delta y_\mathrm{lab}  \Delta p_\mathrm{T} \times (\mathrm{Acc}\times \varepsilon)_{\mathrm{prompt}}(p_\mathrm{T})  \times \mathrm{BR} \times \mathscr{L}_\mathrm{int}},
\label{eq:crosssection}
\end{equation}

where  $N_\mathrm{raw}^{\Lambda^{+}_{\rm c} + \Lambda^{-}_{\rm c}}$ is the raw yield, $f_\mathrm{prompt}$ is the fraction of prompt \Lc in the measured raw yield, BR is the branching ratio, and $\mathscr{L}_\mathrm{int}$ is the integrated luminosity. The factor $2$ accounts for the presence of both particles and antiparticles in the raw yields, and $\Delta y_\mathrm{lab} \Delta p_\mathrm{T}$ accounts for the widths of the rapidity and transverse momentum intervals. For the interval \lowptbin, the measurement of \Lc is performed for $\Delta y_\mathrm{lab} = 1.6$, under the assumption that the cross section per unit rapidity of \Lc baryons does not significantly change between $|y_\mathrm{lab}| < 0.5$ and $|y_\mathrm{lab}| < 0.8$. This has been verified using PYTHIA 8~\cite{sjostrand2008brief} and FONLL~\cite{Cacciari:1998it, fonllcalc1} simulations.
The factor $({\rm Acc} \times{\it \varepsilon})_{\rm prompt}$ is the product of the geometrical acceptance (Acc) and the reconstruction and selection efficiency ($\varepsilon$) for prompt \Lc candidates in the \LctopKzS channel.
The $({\rm Acc} \times{\it \varepsilon})_{\rm prompt}$ corrections were obtained from MC simulations with the same configuration as
those used for the BDT training described above. For both collision systems, the efficiency correction factor was observed to be constant within the interval \lowptbin when computed in narrower \pt intervals. The $({\rm Acc} \times{\it \varepsilon})_{\rm prompt}$ factor is almost constant as a function of rapidity for $|y_\mathrm{lab}|<0.5$, and falls steeply to zero for $|y_\mathrm{lab}|>0.5$.

The fraction of the raw \Lc yield originating from beauty-hadron decays in the selected candidate sample was obtained following the strategy defined in Ref.~\cite{ALICE:2020wla} using: i) the beauty-meson production cross section from FONLL calculations, which is used as a basis for the \pT shape for all beauty-hadron species~\cite{fonllcalc1,fonllcalc2}; ii) the relative abundances of different beauty-hadron species from LHCb measurements in pp collisions~\cite{LHCb:2019fns}; iii) their decay kinematics from PYTHIA 8; and iv) the selection and reconstruction efficiency of \Lc from beauty-hadron decays, which was estimated from MC simulations. The MC samples were generated with a similar configuration as the training samples described in~\secref{sec:methods}, but instead of a charm--anticharm pair, they included a beauty--antibeauty quark pair in each simulated event, with at least one \Lc among the decay products of the resulting beauty hadrons. The efficiency is similar between prompt and feed-down candidates, as there are no tight selections applied on the decay topology of the \Lc baryon. The possible modification of beauty-hadron production in \pPb collisions was included in the feed-down calculation by scaling the beauty-quark production by a nuclear modification factor $R_{\rm pPb}^{\textrm{feed-down}}$. As for previous ALICE measurements of charm hadrons~\cite{ALICE:2016yta,ALICE:2020wla}, the central value was chosen such that the \RpPb of prompt and feed-down \Lc are equal. 
The values of $({\rm Acc} \times{\it \varepsilon})_{\rm prompt}$, $({\rm Acc} \times{\it \varepsilon})_{\textrm{feed-down}}$, and $f_\mathrm{prompt}$ for \lowptbin are listed in~\tabref{tab:corrections} for both collision systems.

\begin{table}[h!t]
    \centering
    \caption{Correction factors $({\rm Acc} \times{\it \varepsilon})_{\rm prompt}$, $({\rm Acc} \times{\it \varepsilon})_{\textrm{feed-down}}$, and $f_\mathrm{prompt}$ in the interval \lowptbin within the measured rapidity regions.} 
    \begin{tabular}{c c c}
    \hline \hline
                            & pp        & \pPb  \\ \hline 
    $({\rm Acc} \times{\it \varepsilon})_{\rm prompt}$    & $\left(6.30 \pm 0.03\right)\%$     & $\left(4.77 \pm 0.02\right)\%$  \\
    $({\rm Acc} \times{\it \varepsilon})_{\textrm{feed-down}}$    & $\left(6.15 \pm 0.03\right)\%$     & $\left(4.71 \pm 0.02\right)\%$  \\
    
    $f_\mathrm{prompt}$ & $\left(98.2^{+0.9}_{-1.5}\right)\% $     &    $\left(98.1^{+0.9}_{-3.7}\right)\% $  \\[0.5ex]
     \hline \hline
    \end{tabular}
    \label{tab:corrections}
\end{table}

\section{Systematic uncertainties} \label{sec:systematics}

The contributions to the systematic uncertainty on the \Lc production cross section in \lowptbin are summarised in \tabref{tab:systematics}.

\begin{table}[t!hb]
    \centering
    \caption{Systematic uncertainties on the \Lc production cross section for pp and \pPb collisions in the \pt interval \lowptbin.} \label{tab:systematics}
    \begin{tabular}{c c c}
    \hline \hline
                            & pp        & \pPb  \\ \hline 
    Raw yield extraction    & $8\%$     & $9\%$  \\
    Selection efficiency    & $9\%$     & $9\%$ \\
    Tracking efficiency     & $4\%$     & $6\%$  \\ 
    Monte Carlo \pt shape   & negl.        & $1\%$ \\
    Feed-down subtraction   & ${}^{+0.9}_{-1.5}\%$ & ${}^{+0.9}_{-3.8}\% $ \\
    Luminosity              & $2.1\%$   & $3.7\%$ \\ 
    Branching ratio  & \multicolumn{2}{c}{$5\%$}
    \\ \hline \hline
    \end{tabular}
\end{table}

The systematic uncertainty on the raw yield extraction was evaluated by repeating the fit to the invariant mass distributions while varying: i) the function used to describe the background, ii) the minimum and maximum of the mass ranges (sidebands) considered for the background fit, iii) the width of the mass peak by $\pm10\%$ compared to the value obtained from MC, and iv) the width of the mass intervals in the invariant mass distribution.
In order to test the sensitivity to the line-shape of the signal, a bin-counting method was used, in which the signal yield was obtained by integrating the invariant-mass distribution after subtracting the background estimated from the fit. The systematic uncertainty was taken as the RMS of the resulting raw-yield distribution, which corresponds to $8\%$ ($9\%$) for the analysis in pp (\pPb) collisions.

The systematic uncertainty on the selection efficiency arises due to possible differences between the real detector resolutions and alignment, and their description in the simulation. This uncertainty was assessed by comparing the production cross sections obtained using different selection criteria. In particular, the selections on the BDT outputs were varied in a range corresponding to a modification of about $30\%$ in the efficiency for both pp and \pPb collisions. The systematic uncertainty was assigned by adding in quadrature the RMS and shift in the mean of the resulting production cross section distribution with respect to the value obtained with the default selections. For both pp and \pPb collisions, this resulted in an uncertainty of $9\%$.  

The tracking efficiency uncertainty was determined by varying the track quality selection criteria and comparing the matching efficiency between the TPC and ITS in data and MC, as described in Ref.~\cite{ALICE:2020wla}. The uncertainties on the individual tracks were propagated to the \Lc candidates according to the decay kinematics, resulting in an uncertainty of 3\% (6\%) in pp (\pPb) collisions. A further contribution was added to account for the imperfect description of the material budget of the detector in the MC simulations, which especially affects the absorption of protons and thus the reconstruction efficiency. This was determined by comparing the corrected yields of charged pions, kaons, and protons using a standard MC production and one with the material budget increased by 10\%, which corresponds to a $2\sigma$ modification based on the estimated systematic uncertainty on the ALICE material budget~\cite{ALICE:2012wos}. The resulting uncertainty on the \Lc yield is 2\% in the interval \lowptbin, leading to an overall tracking efficiency uncertainty of 4\% in pp collisions and 6\% in \pPb collisions.

The possible systematic uncertainty due to the dependence of the efficiencies
on the generated \pt distribution of \Lc
in the simulation was studied (“Monte Carlo \pT shape” in \tabref{tab:systematics}). It was verified that the acceptance and the  reconstruction efficiency do not significantly vary within the \lowptbin interval. Following the same procedure as in Ref.~\cite{ALICE:2020wla}, the efficiencies were evaluated after reweighting the \pt shape of the PYTHIA 8 simulations
to match the \pt spectrum of D mesons from FONLL pQCD calculations~\cite{fonllcalc1,fonllcalc2}, as no FONLL calculations exist for charm baryons. An uncertainty was assigned based on the difference between the nominal and reweighted efficiencies. No significant variation was observed in pp collisions, while a 1\% variation was observed and assigned as systematic uncertainty in \pPb collisions. 

The systematic uncertainty on the feed-down subtraction 
was evaluated by considering the theoretical uncertainties of the beauty-meson production cross section in FONLL~\cite{fonllcalc1,fonllcalc2}, and the variation of the beauty fragmentation function describing the hadronisation $f(\rm b \to \Lb)$  within its uncertainties as measured in Ref.~\cite{LHCb:2019fns}. For \pPb collisions a further consideration is made, varying the ratio of the feed-down and prompt \Lc nuclear modification factors $\RpPb^{\mathrm{feed\textrm{-}down}}/\RpPb^\mathrm{prompt}$ within the range 0.9--3.0. 
The upper bound of this range accounts for recent measurements by LHCb of the nuclear modification of \Lb baryons~\cite{LHCb:2019avm}, where the nuclear modification factor at backward rapidity was found to be consistent with unity. The overall envelope from the variations was considered as the total uncertainty, resulting in ${}^{+0.9}_{-1.5}\%$ in pp collisions and ${}^{+0.9}_{-3.8}\%$ in \pPb collisions.

The production cross section has an additional global normalisation uncertainty due to the integrated luminosity determination. The luminosity uncertainty was determined from van der Meer scans of pp and \pPb collisions at \sqrtsNNfive, and has a value of $2.1\%$ for the pp data sample~\cite{ALICE-PUBLIC-2018-014} and $3.7\%$ for \pPb collisions~\cite{ALICE:2014gvw}. 

The 5\% branching ratio uncertainty for the decay channel $\LctopKs(\to \mathrm{p\pi^+\pi^-})$ is calculated as the quadratic sum of the branching ratio uncertainties for \LctopKs and $\Kzs\to\pi^+\pi^-$~\cite{pdg2022}. This uncertainty is considered as fully correlated between \pt intervals and collision systems.

\section{Results} \label{sec:results}

\begin{figure}[t!hb]
    \centering
    \includegraphics[width=0.65\textwidth]{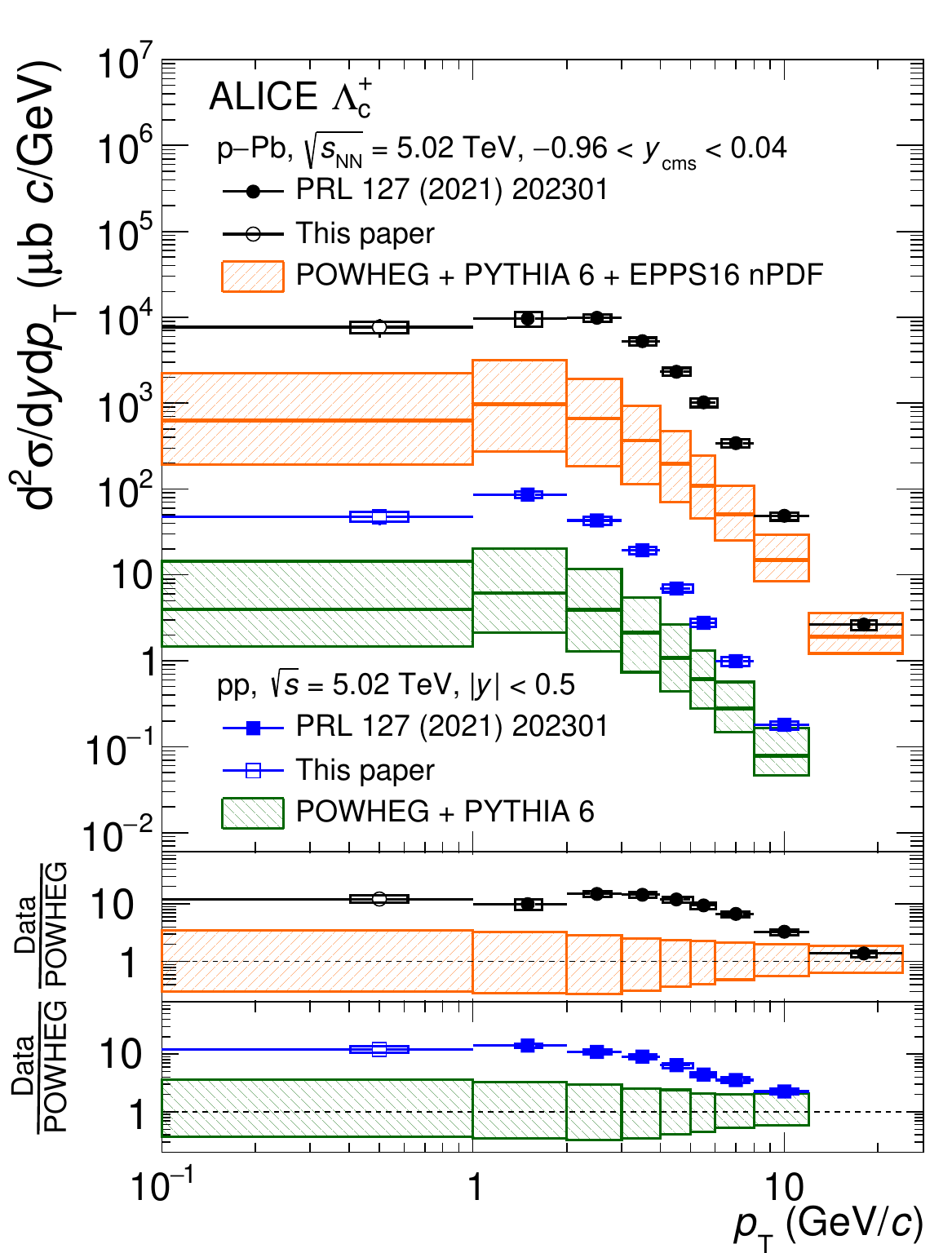} 
    \caption{The \pt-differential \Lc production cross sections in pp and \pPb collisions at \sqrtsNNfive~\cite{ALICE:2020wfu}, including the new measurements in \lowptbin as open markers. The lower panels show the ratios of the measurements to POWHEG+PYTHIA6, with EPPS16 nPDF calculations included for \pPb collisions~\cite{Frixione:2007nw,Skands:2014pea,Eskola:2016oht}.} 
    \label{fig:crosssec}
\end{figure}

The \pt-differential \Lc production cross sections were calculated according to \Eqref{eq:crosssection} and are shown in~\figref{fig:crosssec}, where blue markers are used for pp collisions and black markers for p--Pb collisions. In each collision system, the new result in \lowptbin is shown as an open marker, and the filled markers represent the previous measurements for $\pt>1~\GeVc$ from Refs.~\cite{ALICE:2020wfu,ALICE:2020wla}. 
The \Lc production cross sections are compared with next-to-leading-order (NLO) pQCD calculations obtained with the POWHEG framework~\cite{Frixione:2007nw}, matched with PYTHIA 6~\cite{sjostrand2006pythia} to generate the parton shower and fragmentation, and the CT14NLO parton distribution functions~\cite{Dulat:2015mca}. For \pPb collisions, the nuclear modification of the parton distribution functions is modelled with the EPPS16 nuclear PDF (nPDF) parameterisation~\cite{Eskola:2016oht}.
The nominal factorisation and renormalisation scales, $\mu_{\rm F}$ and $\mu_{\rm R}$, were taken to be equal to the transverse mass of the quark, $\mu_0$ = $\sqrt{m_\mathrm{c}^2+p^2_{\rm T}}$, and the charm-quark mass was set to $m_{\rm c} =1.5$~\GeVcc. The theoretical uncertainties were estimated by varying these scales in the range $0.5\mu_0 < \mu_{\rm R,F} < 2.0\mu_0$, with the constraint $0.5 < \mu_{\rm R}/\mu_{\rm F} <2.0$, as described in Ref.~\cite{fonllcalc1}. For the p--Pb case, the uncertainties on the parton distribution functions and EPPS16 nPDF are not included in the calculation as they are considerably smaller than the scale uncertainties.
In both collision systems the measured \pt-differential production cross section values are significantly underestimated by the POWHEG predictions. In particular, in the interval \lowptbin the model underestimates the measurements by a factor of about 10, similar to what was observed up to $\pt = 3~\GeVc$ in Ref.~\cite{ALICE:2020wla}.

The measured differential production cross sections in \lowptbin are reported  in~\tabref{tab:lowptsigma} and compared with the values from Ref.~\cite{ALICE:2020wla}, where the \lowptbin region was determined from an extrapolation. For both pp and \pPb collisions, the measured values are lower than the extrapolated ones and have smaller overall uncertainties, but remain within 1$\sigma$ when considering the combined measurement and extrapolation uncertainties.
The previously computed extrapolated production cross section in pp collisions was based on PYTHIA 8 predictions with specific tunes implementing colour-reconnection mechanisms beyond the leading-colour approximation, and the extrapolation uncertainty was assigned by taking the envelope of the different tunes. In \pPb collisions, the extrapolation was performed by multiplying the extrapolated regions of the production cross section in pp collisions by i) the Pb mass number, ii) a correction factor to account for the different rapidity intervals covered in pp and \pPb collisions, and iii) a factor based on an assumption on the nuclear modification factor \RpPb. The central value was calculated using $\RpPb= 0.5$ and the extrapolation uncertainty was estimated by varying this hypothesis in the range $0.35 < \RpPb < 0.8$~\cite{ALICE:2020wla}.

\begin{table}[t]
    \centering
    \caption{The \Lc production cross sections at \lowptbin in pp collisions for $\left|y\right| < 0.5$ and \pPb collisions for $-0.96<y_\mathrm{cms}<0.04$, at \sqrtsNNfive. The left values are the new measurements from this article, and the right ones are the previously extrapolated values from Ref.~\cite{ALICE:2020wla}.} 
    \begin{tabular}{c c c}
    \hline \hline
      \multicolumn{3}{c}{$\mathrm{d}^2\sigma/\mathrm{d}p_\mathrm{T}\mathrm{d}y\,\,(0<\pt<1~\GeVc)$} \\ \hline
                           & measured        & extrapolated~\cite{ALICE:2020wla}  \\ \hline \rule{0pt}{\normalbaselineskip}
    pp (\mub (\GeVc)$^{-1}$)  &  $47.9 \pm 10.4~(\mathrm{stat.})\pm 6.1~(\mathrm{syst.}) \pm 1.0~(\mathrm{lumi.})$  &   $68.5^{+11.9}_{-15.9}~(\mathrm{extr.}) $\\ \rule{0pt}{\normalbaselineskip}
    \pPb (\mb (\GeVc)$^{-1}$) & $ 7.7 \pm 1.9~(\mathrm{stat.})\pm 1.1~(\mathrm{syst.}) \pm 0.3~(\mathrm{lumi.})$ & $8.5^{+5.1}_{-2.6}~(\mathrm{extr.})$ 
  \\[0.5ex]
     \hline \hline
    \end{tabular}
    \label{tab:lowptsigma}
\end{table}

The production cross section measurement in the interval \lowptbin allows the \pt-integrated production cross section to be calculated without the need for a model-dependent extrapolation, which in the previous publication~\cite{ALICE:2020wla} accounted for about 30\% (20\%) of the total \Lc production cross section in pp (p--Pb) collisions. 
The rapidity-differential production cross sections for \lowptbin were summed with the values measured for the region $1<\pt < 12~(24)~\GeVc$ for pp (\pPb) collisions in Ref.~\cite{ALICE:2020wla} to obtain the integrated cross section. No extrapolation towards higher \pt is performed in either system, as the contribution to the \pt-integrated production cross section is negligible ($<0.1\%$) for the reported level of precision. The systematic uncertainties due to the raw-yield extraction were propagated as uncorrelated between \pt intervals, and all other sources were considered as fully correlated. The resulting \pt-integrated prompt \Lc production cross sections in the two collision systems are reported in~\tabref{tab:integSigma}, and compared with the values published in Ref.~\cite{ALICE:2020wla} based on the \pt extrapolation described above.

\begin{table}[h!t]
    \centering
    \caption{The \pt-integrated production cross sections for prompt \Lc baryons in pp collisions for $|y|<0.5$ and \pPb collisions for $-0.96<y_\mathrm{cms}<0.04$, at \sqrtsNNfive. The first two rows correspond to the measured values over the full \pt range, and the last two rows to the previously extrapolated results from Ref.~\cite{ALICE:2020wla}.} 
    \begin{tabular}{c c }
    \hline \hline
 & \multicolumn{1}{c}{$ \mathrm{d} \sigma^{\Lc}/\mathrm{d} y$ }\\ \hline
pp, measured (\mub) &  $208 \pm 15~(\mathrm{stat.}) \pm 15~(\mathrm{syst.}) \pm 4~(\mathrm{lumi.})$ \\
\pPb, measured (\mb) & $36.9 \pm 3.3~(\mathrm{stat.}) \pm 4.5~(\mathrm{syst.}) \pm 1.4~(\mathrm{lumi.})$ \\
pp, extrapolated (\mub)~\cite{ALICE:2020wla} & $230 \pm 16~(\mathrm{stat.}) \pm 20~(\mathrm{syst.}) \pm 5~(\mathrm{lumi.})^{+5}_{-10}~(\mathrm{extr.})$ \\
\pPb, extrapolated (\mb)~\cite{ALICE:2020wla} & $36.2 \pm 2.5~(\mathrm{stat.}) \pm 4.5~(\mathrm{syst.})\pm1.3~(\mathrm{lumi.})^{+4.4}_{-2.7}~(\mathrm{extr.})$
\\[0.5ex]
     \hline \hline
    \end{tabular}
    \label{tab:integSigma}
\end{table}

The new measurement in the \lowptbin interval in pp collisions results in a reduction of the \pt-integrated \Lc production cross section by about 10\% with respect to the previous published results, but the two values remain compatible in terms of the combined statistical and systematic uncertainties. In p--Pb collisions the \pt-integrated \Lc production cross section is also compatible with the previous measurement~\cite{ALICE:2020wla}.

\begin{figure}[t!]
    \centering
    \includegraphics[width=0.6\textwidth]{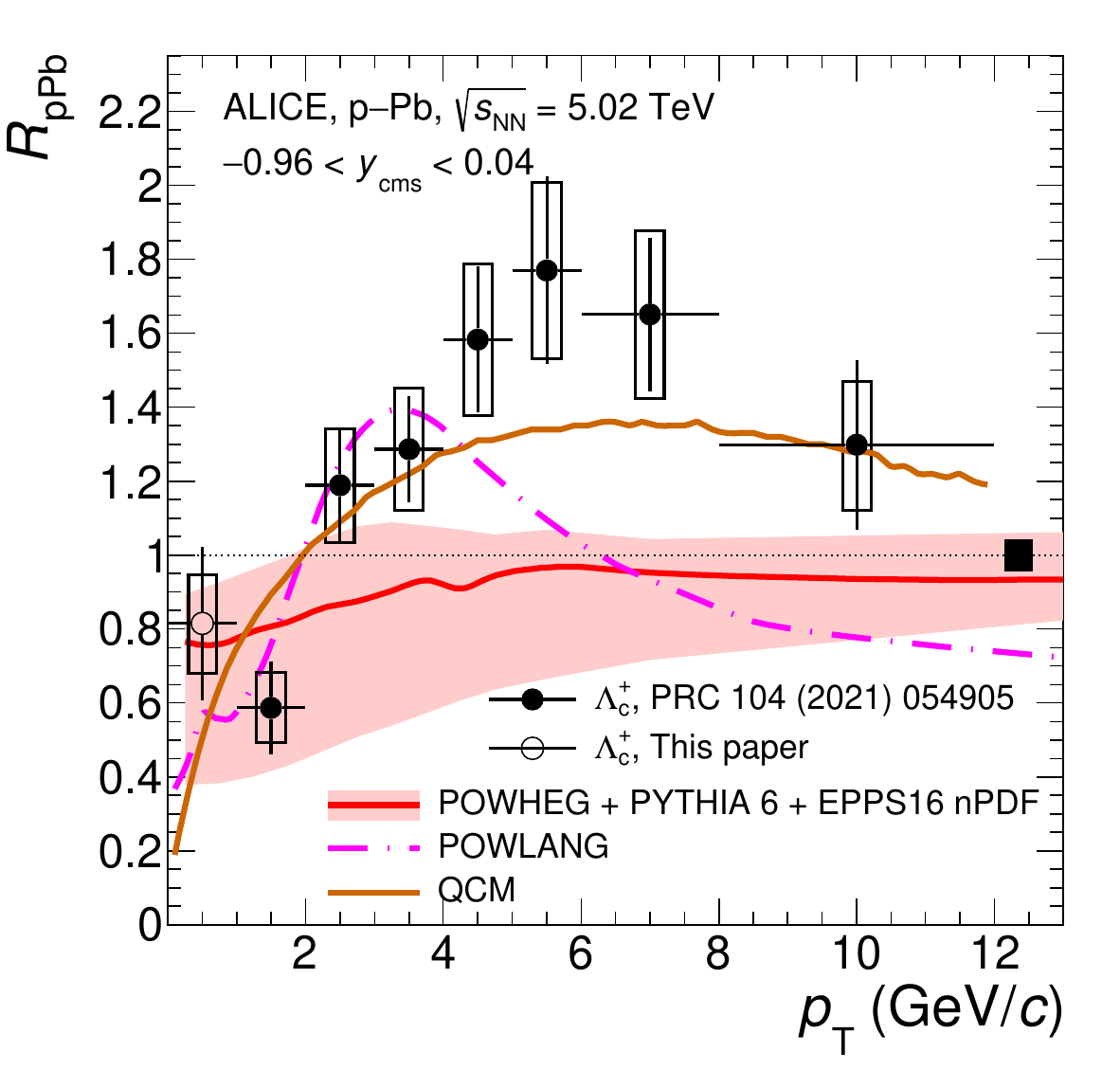} 
    
    \caption{Nuclear modification factor \RpPb of prompt \Lc baryons in \pPb collisions at \sqrtsNNfive as a function of \pt, compared with model calculations~\cite{Frixione:2007nw,Eskola:2016oht,Beraudo:2015wsd,Li:2017zuj}.}
    \label{fig:RpPb}
\end{figure}

In order to compare the spectral shapes in the two different collision systems at the same energy, the nuclear modification factor \RpPb, which is the ratio between the \Lc production cross sections in \pPb and pp collisions, scaled by the nuclear mass number $A=208$ and corrected to account for the shift in rapidity between pp and \pPb collisions using FONLL~\cite{fonllcalc1}, is calculated. The systematic uncertainties on the branching ratio and beauty feed-down  are considered as fully correlated between the two collision systems, and all other systematic uncertainties as uncorrelated. This is shown as a function of \pt in~\figref{fig:RpPb}. The \RpPb in \lowptbin is consistent with unity within the uncertainties, and is also consistent with the decreasing trend towards low \pt within $0<\pt<6~\GeVc$ that was previously observed in Ref.~\cite{ALICE:2020wla}. 
 The results are compared with the POWHEG+PYTHIA6~\cite{Frixione:2007nw,Eskola:2016oht} and POWLANG~\cite{Beraudo:2015wsd} models, as well as the QCM model~\cite{Li:2017zuj}.
In the QCM model, the charm quark is combined with a co-moving light antiquark or with two co-moving quarks to form a charm meson or baryon.
For light-flavour (u, d, and s) quarks, the momentum distribution is obtained by fitting the data of hadronic \pt spectra using the quark coalescence formulas of QCM and parameterising the hadron and quark spectra with a $\rm L\acute{e}vy$-Tsallis function, as explained in Ref.~\cite{Gou:2017foe}.
A free parameter, $ R^{(\rm c)}_{\rm B/M}$, characterises the relative production of single-charm baryons to single-charm mesons. This value is set to 0.425, which is tuned to reproduce the $\Lambda^+_{\rm c}/{\rm D^0}$ ratio measured by ALICE in pp collisions at \sqrtsseven~\cite{ALICE:2017thy}. The relative abundances of the different charm-baryon species are determined by thermal weights from the statistical hadronisation approach~\cite{Andronic:2009sv}.
The POWHEG+PYTHIA6 pQCD event generator, which is coupled with the EPPS16 nPDF set for \pPb collisions, predicts a central \RpPb value that is below unity for all \pt and constant for $\pt>4~\GeVc$, but consistent with unity within the uncertainties. It should be noted that the uncertainties on this calculation come solely from the EPPS16 nPDF parametrisation, as the uncertainties related to the pQCD scales in the POWHEG+PYTHIA6 calculation cancel out in the ratio between \pPb and pp collisions. While the model is in fair agreement with the measurements for $\pt<3~\GeVc$, it does not describe the increase above unity in the region $4<\pt<8~\GeVc$. Similarly, the POWLANG calculations are peaked in the region $2<\pt<4~\GeVc$, but are at tension with the data for $\pt>4~\GeVc$. In the case of POWLANG, the \RpPb is the result of the transport of charm quarks through an expanding quark--gluon plasma, which is assumed to be formed in \pPb collisions and affects the \pT distributions of charm hadrons. However, the calculated value is identical for all charm-hadron species as it does not consider any modifications of the relative hadron abundances due to quark coalescence. The QCM model, which does not include any nPDF or cold nuclear matter effects, gives the closest description of the measurement over the full measured \pt range.

The \pt-integrated \RpPb of prompt \Lc baryons was calculated from the \pt-integrated production cross sections measured in p--Pb and pp collisions, and is reported in~\tabref{tab:integRaa}.  The value is consistent with the atomic mass number scaling of the \Lc production cross section in pp collisions (i.e.~$\RpPb=1$), within 1.1$\sigma$ of the combined statistical and systematic uncertainties.
The \pt-integrated production cross section in pp collisions from~\tabref{tab:integSigma} is also used to compute the \Raa of prompt \Lc baryons from the \pt-integrated corrected yields in central (0--10\%) and semi-central (30--50\%) Pb--Pb collisions at \sqrtsNNfive reported in Ref.~\cite{ALICE:2021bib}. These values are also reported in~\tabref{tab:integRaa}. The extrapolation uncertainties on the \PbPb nuclear modification factors arise due to the extrapolation of the \PbPb \Lc-baryon yields down to \pt$=0$, which was performed by estimating the \LcD ratio in \lowptbin with model calculations~\cite{Plumari:2017ntm,He:2019vgs,Andronic:2021erx,PhysRevC.48.2462} and multiplying it by the measured \Dzero-meson yield~\cite{ALICE:2021rxa}. The uncertainty was determined from the variation of the resulting \Lc yield with different model calculations.

\begin{table}[h!t]
    \centering
    \caption{The \pt-integrated nuclear modification factors \RpPb and \Raa of prompt \Lc baryons in \pPb and \PbPb collisions at \sqrtsNNfive. The \PbPb results are derived from the integrated yields published in Ref.~\cite{ALICE:2021bib}. The percentile ranges in the first column represent the centrality ranges considered for \PbPb collisions.} 
    \begin{tabular}{c c }
    \hline \hline
    & \Lc nuclear modification factor \\ \hline
\pPb & $ 0.85 \pm 0.09 ~(\mathrm{stat.}) \pm 0.11~(\mathrm{syst.})$ \\ 
\PbPb (0--10\%) &   $0.68 \pm 0.10~(\mathrm{stat.})\pm 0.10~(\mathrm{syst.}) ^{+0.10} _{-0.06}~(\mathrm{extr.})$ \\[0.5ex]
\PbPb (30--50\%)  & $0.86 \pm 0.13 ~(\mathrm{stat.})\pm 0.13~(\mathrm{syst.}) ^{+0.09} _{-0.06}~(\mathrm{extr.})$
\\[0.5ex]
     \hline \hline
    \end{tabular}
    \label{tab:integRaa}
\end{table}

\begin{figure}[h!tb]
    \centering
    \includegraphics[width=0.7\textwidth]{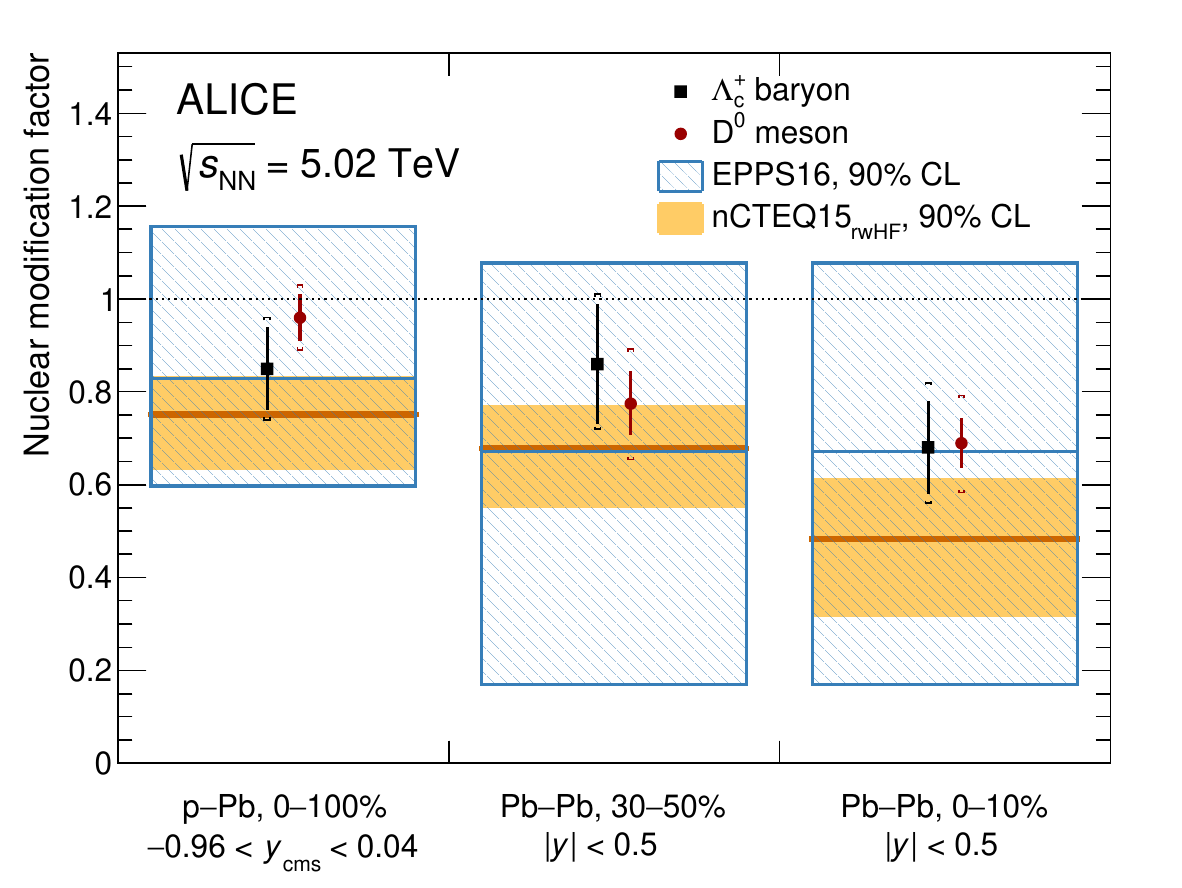} 
    \caption{The \pt-integrated nuclear modification factors of prompt \Lc baryons and \Dzero mesons measured in \pPb and Pb–Pb collisions at \sqrtsNNfive~\cite{ALICE:2021bib,ALICE:2021rxa}. Statistical (bars) and systematic and extrapolation (brackets) uncertainties are shown. The measurements are compared with calculations from the theoretical models nCTEQ15~\cite{Kusina:2017gkz,Kusina:2020dki,Kovarik:2015cma} and EPPS16~\cite{Eskola:2016oht} that include only initial-state effects. The uncertainty bands on the models represent the 90\% confidence level.}
    \label{fig:RAA}
\end{figure}

Figure~\ref{fig:RAA} shows the \pt-integrated nuclear modification factors for \Lc baryons in \pPb and \PbPb collisions, compared with those measured for $\rm D^{0}$ mesons in Ref.~\cite{ALICE:2021rxa}. 
The \pt-integrated \Raa of \Lc is $1.8\sigma$ below unity in 0--10\% central collisions, indicating a suppression of the \Lc-baryon yield in Pb--Pb collisions with respect to the binary-scaled pp reference due to shadowing and possible modifications in the hadronisation mechanism. In the 30--50\% centrality interval, the \pt-integrated \Lc \Raa is compatible with unity within the uncertainties. The \pt-integrated \Lc \RpPb is closer to unity than the \RAA in central \PbPb collisions, as expected from the smaller shadowing effects in p--Pb compared to Pb--Pb collisions, where the nucleons of both the projectile and the target nuclei are involved. In all three collision systems, the nuclear modification factors for \Lc and \Dzero are consistent with one another, indicating that there is no significant enhancement of the overall production of charm baryons compared to charm mesons in heavy-ion collisions.
The integrated \Raa and \RpPb are also compared with perturbative QCD calculations including only initial-state effects modeled using two different sets of nuclear PDFs, namely a Bayesian-reweighted version~\cite{Kusina:2017gkz,Kusina:2020dki} of nCTEQ15~\cite{Kovarik:2015cma} and EPPS16~\cite{Eskola:2016oht}. The calculations with EPPS16 do not include the dependence of the shadowing on the impact parameter of the Pb--Pb collision, and therefore they are identical in the central and semi-central event classes. The predictions with nCTEQ15 are obtained by applying a Bayesian reweighting of the nuclear PDFs, which is constrained by measurements of heavy-flavour hadron production in \pPb collisions at the LHC~\cite{Kusina:2017gkz}, and are labelled as nCTEQ15$_\mathrm{rwHF}$ in Fig.~\ref{fig:RAA}. The uncertainty bands for both calculations represent the 90\% confidence level regions. In the reweighted nCTEQ15 case they are determined by considering three different factorisation scales in addition to the PDF uncertainties. The measured \Raa and \RpPb values are within the upper edge of the nCTEQ15 uncertainty band. These data provide an important input for testing the assumptions of nPDFs in theoretical calculations.

\begin{figure}[h!tb]
    \centering
    \includegraphics[width=0.49\textwidth]{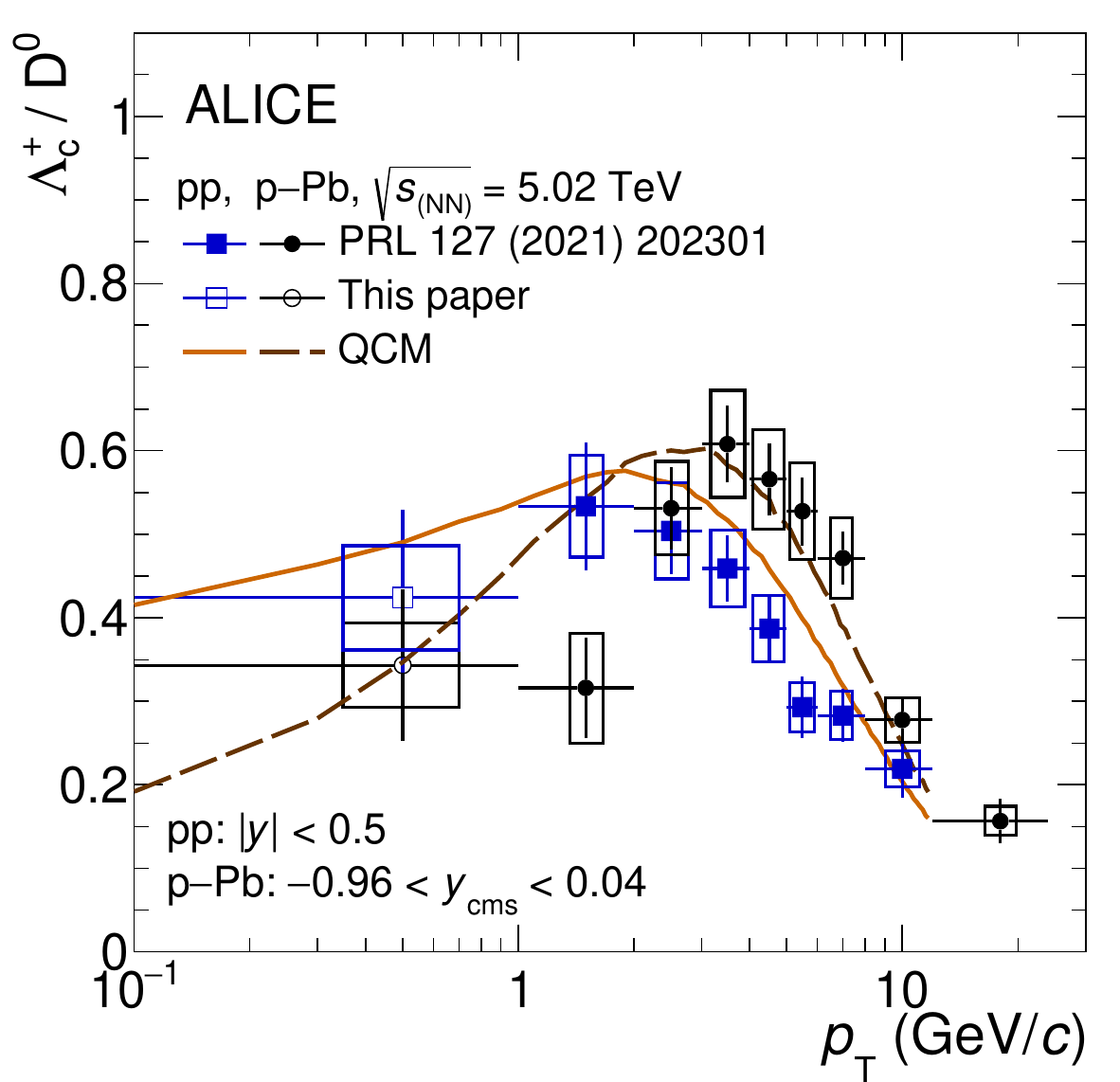}
    \includegraphics[width=0.49\textwidth]{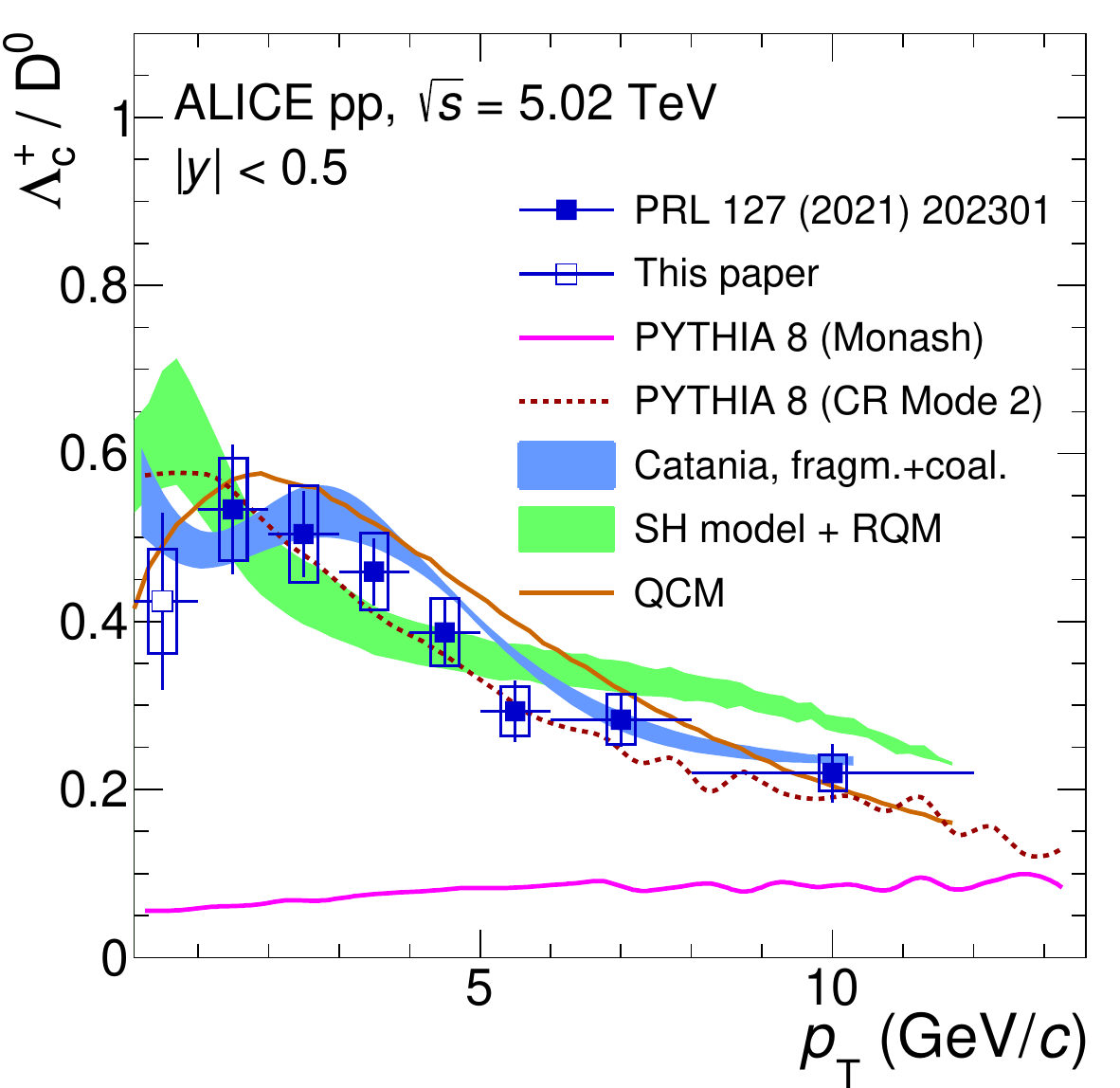} 
    \caption{Left: \LcD ratio in pp and \pPb collisions as a function of \pt, compared with the QCM model~\cite{Li:2017zuj,Li:2021nhq}. Right: \LcD ratio as a function of \pt in pp collisions at \sqrtsfive, including comparisons with models~\cite{He:2019tik,Christiansen:2015yqa,Skands:2014pea,Minissale:2020bif,Li:2021nhq}} 
    \label{fig:LcDWithModels}  \label{fig:LcD}
\end{figure}

The \LcD baryon-to-meson yield ratio is used to further examine differences in the charm-quark hadronisation into baryons and mesons that may arise due to the differing numbers of constituent quarks. The results in pp and \pPb collisions are shown in the left panel of~\figref{fig:LcD}. The \pt-differential \Dz production cross section in \lowptbin was taken from Ref.~\cite{ALICE:2021mgk} for pp collisions and from Ref.~\cite{ALICE:2019fhe} for \pPb collisions. In the calculation of the baryon-to-meson ratio, the uncertainties related to the tracking efficiency, luminosity, and beauty feed-down were treated as fully correlated between the two species, and all other uncertainty contributions were considered to be uncorrelated.
The \LcD yield ratio in \lowptbin in both pp and \pPb collisions indicates a decreasing trend with respect to the intermediate \pt region, albeit with large uncertainties. Within uncertainties, the \LcD ratios are consistent between pp and \pPb collisions. The distribution has a maximum in the region $1<\pt<3$ ($3<\pt<5$)~\GeVc in pp (\pPb) collisions. The shift of the peak towards higher \pT in \pPb collisions could be attributed to a contribution of collective effects, e.g.~radial flow. Similar collective effects have been observed for light- and heavy-flavour hadrons in \pPb collisions at the LHC~\cite{CMS:2014und,ATLAS:2016yzd,ALICE:2018gyx}. Such a contribution would be consistent with previous observations for the light-flavour $\Lambda/\Kzs$ baryon-to-meson ratio~\cite{ALICE:2013wgn}. The results are also compared with the QCM model~\cite{Song:2018tpv,Li:2021nhq} which describes the magnitude of the \LcD ratio well for $0<\pt<12~\GeVc$ in both collision systems, as well as predicting a shift of the peak towards higher \pt, resulting from a hardening of the \Lc spectrum in \pPb collisions.

The modification of the \Lc-baryon production spectrum in \pPb collisions is confirmed by computing the mean transverse momentum, \meanpt. This was calculated in each collision system following the same prescription as in Ref.~\cite{ALICE:2016yta}, with the central value derived from a power-law fit to the \pt spectrum. The resulting values are summarised in~\tabref{tab:meanpt} and compared with the values obtained for \Dzero mesons in Ref.~\cite{ALICE:2019fhe}. 
The \meanpt value for \Lc baryons is significantly higher in \pPb collisions than in pp collisions, by $3.7\sigma$ considering the combined statistical and systematic uncertainties. This is in contrast with the results for \Dzero mesons, for which the \meanpt is seen to be fully consistent between the two collision systems.

\begin{table}[h!t]
    \centering
    \caption{Mean transverse momentum values for \Dzero mesons~\cite{ALICE:2019fhe} and \Lc baryons  in pp and \pPb collisions at \sqrtsNNfive.} 
    \begin{tabular}{c c c}
    \hline \hline 
\multicolumn{3}{c}{    \meanpt (\GeVc)} \\\hline
                            & pp        & \pPb  \\ \hline \rule{0pt}{\normalbaselineskip}
    \Dzero & $2.06 \pm 0.03~\mathrm{(stat.)} \pm 0.03~\mathrm{(syst.)}$   & $2.07 \pm 0.02~\mathrm{(stat.)}\pm0.04~\mathrm{(syst.)}$ \\ \rule{0pt}{\normalbaselineskip}
    \Lc & $1.86 \pm 0.06~ \mathrm{ (stat.)}\pm 0.03 ~\mathrm{ (syst.)}$ & $2.29 \pm 0.06 ~\mathrm{ (stat.)}\pm 0.06~\mathrm{ (syst.)}$ \\[0.5ex] 
     \hline \hline
    \end{tabular}
    \label{tab:meanpt}
\end{table}

The right panel of Fig.~\ref{fig:LcD} shows the \LcD yield ratio in pp collisions as a function of \pt compared with model calculations in which different hadronisation processes are implemented.
The Monash tune of PYTHIA 8~\cite{Skands:2014pea}, which implements fragmentation processes tuned on charm-hadron production measurements in \ee collisions, predicts an integrated value of about 0.1 for the \LcD ratio, with a mild \pT dependence. This significantly underpredicts the data, as already seen in Refs.~\cite{ALICE:2020wla,ALICE:2020wfu}, with a difference of approximately a factor 8 between the data and model in the interval \lowptbin. Model calculations including processes that
enhance baryon production, like PYTHIA
8 including colour reconnection beyond the leading-colour approximation~\cite{Christiansen:2015yqa}, SHM+RQM~\cite{He:2019tik}, QCM~\cite{Li:2021nhq}, and Catania~\cite{Minissale:2020bif} are also shown. Hadronisation in PYTHIA 8 is built on the Lund string fragmentation model~\cite{ANDERSSON198331}, where quarks and gluons connected by colour strings fragment into hadrons, and colour reconnection allows for partons created in the collision to interact via colour strings. The tune with colour-reconnection topologies beyond the leading-colour approximation includes so-called ``junctions'' that fragment into baryons and lead to increased baryon production with respect to the Monash tune. The statistical hadronisation model includes additional excited charm-baryon states that have not yet been observed but are predicted by the Relativistic Quark Model~\cite{Ebert:2011kk}. These additional states decay strongly to \Lc baryons, thereby contributing to
the prompt \Lc spectrum. The SHM+RQM predictions include a source of uncertainty related to the branching ratios of the excited baryon states into \Lc final states, which is estimated by varying the branching ratios between 50\% and 100\%.
The Catania model assumes that a colour-deconfined state of matter is formed in pp collisions, and hadronisation can occur via quark coalescence in addition to fragmentation. Coalescence is implemented through the Wigner formalism, where a blast-wave model
is used to determine the \pt spectrum of light quarks, and FONLL pQCD calculations are used for heavy quarks. Hadronisation via coalescence is predicted to dominate at low \pt, while fragmentation dominates at high \pt.
All of these models qualitatively reproduce the data. The QCM model predicts a maximum in the region $1<\pt<3~\GeVc$, while the other models tend to predict a continuous increase of the \LcD yield ratio towards low \pt, reaching a value of about 0.6 at $\pt = 0$. This trend might highlight some tension with the data in the interval \lowptbin, since the data hint to a decrease of the \LcD yield ratio, though a more precise measurement is needed to reach a firm conclusion.

\begin{table}[h!t]
    \centering
    \caption{The \pt-integrated \LcD yield ratios in pp and \pPb collisions at \sqrtsNNfive.} 
    \begin{tabular}{c c}
    \hline \hline 
&   \LcD  \\\hline
 pp     & $  0.47 \pm 0.04~(\mathrm{stat.})\pm0.04~(\mathrm{syst.})$ \\
  \pPb  & $ 0.42 \pm 0.04~(\mathrm{stat.}) \pm 0.06~(\mathrm{syst.})$   \\[0.5ex]   
     \hline \hline
    \end{tabular}
    \label{tab:integLcD}
\end{table}

The \pt-integrated \LcD yield ratios in pp and \pPb collisions are presented in~\tabref{tab:integLcD}.
These are consistent with each other within 1$\sigma$ of the combined statistical and systematic uncertainties, indicating no modification of the overall hadronisation fractions between pp and \pPb collisions despite the modification of the \meanpt. A similar effect was observed for \Lc baryons measured as a function of charged-particle multiplicity in pp collisions at \sqrtsthirt~\cite{ALICE:2021npz}, where the \pt-integrated \LcD ratio was found to be independent of multiplicity despite a significant alteration of the \pt-dependent spectrum. This could indicate a common mechanism that alters the \pt distribution of charm baryons for \pPb and high-multiplicity pp collisions while leaving the integrated relative abundance of baryons and mesons consistent with lower-multiplicity pp collisions.

\section{Summary} \label{sec:summary}

The first measurements of the production of prompt \Lc baryons in the transverse momentum interval \lowptbin in pp ($|y| < 0.5$) and in p--Pb ($-0.96<y_\mathrm{cms}<0.04$) collisions at \sqrtsNNfive with the ALICE detector at the LHC are reported, removing the model dependence affecting the previous results for the \pt-integrated \Lc yields~\cite{ALICE:2020wla}. The analysis was performed using the decay channel \LctopKs. The \Lc production cross section in the interval \lowptbin was measured to be larger than predictions given by pQCD-based calculations in both pp and \pPb collisions. The uncertainties on the two measurements are smaller than the theoretical uncertainties on the previously extrapolated values~\cite{ALICE:2020wla}. The \pt-differential \RpPb was measured in \lowptbin and found to be consistent with unity within the uncertainties, and also with a decreasing trend towards low \pt in $0<\pt<6$~\GeVc. However, the current precision of the measurement is not enough to draw firm conclusions on the role of cold nuclear matter effects and on the possible presence of collective effects, like the radial flow, which are observed in heavy-ion collisions. In addition, the \pt-integrated \RpPb and \Raa of prompt \Lc baryons were obtained and compared with those of \Dzero mesons at the same centre-of-mass energy, showing compatibility between the nuclear modification factors of the two charm hadron species. The results are consistent with calculations that consider nuclear modification of the PDFs.

The \LcD yield ratio in \lowptbin in both pp and \pPb collisions indicates a decreasing trend with respect to the intermediate \pt region, albeit with large uncertainties. The PYTHIA 8 event generator with the Monash tune, which incorporates fragmentation parameters from \ee collisions, significantly underestimates the \LcD yield ratio. The data are qualitatively reproduced by models that predict an enhancement of baryon production by various mechanisms, including colour reconnection beyond the leading-colour approximation, feed-down from unobserved resonant charm-baryon states, or quark coalescence (recombination).
The quark (re)combination model also describes the shift of the peak in the \LcD ratio between pp and \pPb collisions. The hardening of the \pt spectrum of \Lc baryons is confirmed by calculating the \meanpt, resulting in a $3.7\sigma$ modification between pp and \pPb collisions. 
The measurement of the \Lc baryon in the interval \lowptbin in pp and p--Pb collisions and the \pt-integrated results are crucial for providing further insight into charm-quark hadronisation in pp and \pPb collisions, and for the investigation of cold nuclear matter effects in \pPb collisions. More precise measurements are expected to be performed during Runs 3 and 4 of the LHC thanks to the upgraded ALICE detector~\cite{ALICE:2012dtf}.

%%%%%%%%%%%%%%%%%%%%%%%%%%%%%%%%
% end main text 
%%%%%%%%%%%%%%%%%%%%%%%%%%%%%%%%

%%%%% acknowledgements - handled by EB chairs 
\newenvironment{acknowledgement}{\relax}{\relax}
\begin{acknowledgement}
\section*{Acknowledgements}
% add specific acknowledgements here 
% ...but please don't remove the line below: funding agencies
% will be acknowledged with a custom tex file handled by EB chairs after Collab Round 2
% Version: 2022-11-09

The ALICE Collaboration would like to thank all its engineers and technicians for their invaluable contributions to the construction of the experiment and the CERN accelerator teams for the outstanding performance of the LHC complex.
The ALICE Collaboration gratefully acknowledges the resources and support provided by all Grid centres and the Worldwide LHC Computing Grid (WLCG) collaboration.
The ALICE Collaboration acknowledges the following funding agencies for their support in building and running the ALICE detector:
A. I. Alikhanyan National Science Laboratory (Yerevan Physics Institute) Foundation (ANSL), State Committee of Science and World Federation of Scientists (WFS), Armenia;
Austrian Academy of Sciences, Austrian Science Fund (FWF): [M 2467-N36] and Nationalstiftung f\"{u}r Forschung, Technologie und Entwicklung, Austria;
Ministry of Communications and High Technologies, National Nuclear Research Center, Azerbaijan;
Conselho Nacional de Desenvolvimento Cient\'{\i}fico e Tecnol\'{o}gico (CNPq), Financiadora de Estudos e Projetos (Finep), Funda\c{c}\~{a}o de Amparo \`{a} Pesquisa do Estado de S\~{a}o Paulo (FAPESP) and Universidade Federal do Rio Grande do Sul (UFRGS), Brazil;
Bulgarian Ministry of Education and Science, within the National Roadmap for Research Infrastructures 2020¿2027 (object CERN), Bulgaria;
Ministry of Education of China (MOEC) , Ministry of Science \& Technology of China (MSTC) and National Natural Science Foundation of China (NSFC), China;
Ministry of Science and Education and Croatian Science Foundation, Croatia;
Centro de Aplicaciones Tecnol\'{o}gicas y Desarrollo Nuclear (CEADEN), Cubaenerg\'{\i}a, Cuba;
Ministry of Education, Youth and Sports of the Czech Republic, Czech Republic;
The Danish Council for Independent Research | Natural Sciences, the VILLUM FONDEN and Danish National Research Foundation (DNRF), Denmark;
Helsinki Institute of Physics (HIP), Finland;
Commissariat \`{a} l'Energie Atomique (CEA) and Institut National de Physique Nucl\'{e}aire et de Physique des Particules (IN2P3) and Centre National de la Recherche Scientifique (CNRS), France;
Bundesministerium f\"{u}r Bildung und Forschung (BMBF) and GSI Helmholtzzentrum f\"{u}r Schwerionenforschung GmbH, Germany;
General Secretariat for Research and Technology, Ministry of Education, Research and Religions, Greece;
National Research, Development and Innovation Office, Hungary;
Department of Atomic Energy Government of India (DAE), Department of Science and Technology, Government of India (DST), University Grants Commission, Government of India (UGC) and Council of Scientific and Industrial Research (CSIR), India;
National Research and Innovation Agency - BRIN, Indonesia;
Istituto Nazionale di Fisica Nucleare (INFN), Italy;
Japanese Ministry of Education, Culture, Sports, Science and Technology (MEXT) and Japan Society for the Promotion of Science (JSPS) KAKENHI, Japan;
Consejo Nacional de Ciencia (CONACYT) y Tecnolog\'{i}a, through Fondo de Cooperaci\'{o}n Internacional en Ciencia y Tecnolog\'{i}a (FONCICYT) and Direcci\'{o}n General de Asuntos del Personal Academico (DGAPA), Mexico;
Nederlandse Organisatie voor Wetenschappelijk Onderzoek (NWO), Netherlands;
The Research Council of Norway, Norway;
Commission on Science and Technology for Sustainable Development in the South (COMSATS), Pakistan;
Pontificia Universidad Cat\'{o}lica del Per\'{u}, Peru;
Ministry of Education and Science, National Science Centre and WUT ID-UB, Poland;
Korea Institute of Science and Technology Information and National Research Foundation of Korea (NRF), Republic of Korea;
Ministry of Education and Scientific Research, Institute of Atomic Physics, Ministry of Research and Innovation and Institute of Atomic Physics and University Politehnica of Bucharest, Romania;
Ministry of Education, Science, Research and Sport of the Slovak Republic, Slovakia;
National Research Foundation of South Africa, South Africa;
Swedish Research Council (VR) and Knut \& Alice Wallenberg Foundation (KAW), Sweden;
European Organization for Nuclear Research, Switzerland;
Suranaree University of Technology (SUT), National Science and Technology Development Agency (NSTDA), Thailand Science Research and Innovation (TSRI) and National Science, Research and Innovation Fund (NSRF), Thailand;
Turkish Energy, Nuclear and Mineral Research Agency (TENMAK), Turkey;
National Academy of  Sciences of Ukraine, Ukraine;
Science and Technology Facilities Council (STFC), United Kingdom;
National Science Foundation of the United States of America (NSF) and United States Department of Energy, Office of Nuclear Physics (DOE NP), United States of America.
In addition, individual groups or members have received support from:
Marie Sk\l{}odowska Curie, European Research Council, Strong 2020 - Horizon 2020 (grant nos. 950692, 824093, 896850), European Union;
Academy of Finland (Center of Excellence in Quark Matter) (grant nos. 346327, 346328), Finland;
Programa de Apoyos para la Superaci\'{o}n del Personal Acad\'{e}mico, UNAM, Mexico.

\end{acknowledgement}

%%%%%%%% Bibliography 
\bibliographystyle{utphys}   % Remember we use title in the biblio
\bibliography{bibliography}
%\input {bibliography.tex}  

%%%%%%%%%%%%%%%%%%%%%%%%%%%%%%%%
% Appendices: yours (if any) + authorlist
%%%%%%%%%%%%%%%%%%%%%%%%%%%%%%%%
\newpage
\appendix

%
%\input{} % put your appendices here (if any)
%

%%%%% Authorlist - please do not touch: handled by EB chairs 
\section{The ALICE Collaboration}
\label{app:collab}
% ALICE Collaboration author list for 2022-11-09
\begin{flushleft} 
\small

S.~Acharya\,\orcidlink{0000-0002-9213-5329}\,$^{\rm 125}$, 
D.~Adamov\'{a}\,\orcidlink{0000-0002-0504-7428}\,$^{\rm 86}$, 
A.~Adler$^{\rm 69}$, 
G.~Aglieri Rinella\,\orcidlink{0000-0002-9611-3696}\,$^{\rm 32}$, 
M.~Agnello\,\orcidlink{0000-0002-0760-5075}\,$^{\rm 29}$, 
N.~Agrawal\,\orcidlink{0000-0003-0348-9836}\,$^{\rm 50}$, 
Z.~Ahammed\,\orcidlink{0000-0001-5241-7412}\,$^{\rm 132}$, 
S.~Ahmad\,\orcidlink{0000-0003-0497-5705}\,$^{\rm 15}$, 
S.U.~Ahn\,\orcidlink{0000-0001-8847-489X}\,$^{\rm 70}$, 
I.~Ahuja\,\orcidlink{0000-0002-4417-1392}\,$^{\rm 37}$
% ...

\section*{Affiliation Notes}

$^{\rm I}$ Deceased\\
$^{\rm II}$ Also at: Max-Planck-Institut f\"{u}r Physik, Munich, Germany\\
$^{\rm III}$ Also at: Italian National Agency for New Technologies, Energy and Sustainable Economic Development (ENEA), Bologna, Italy\\
$^{\rm IV}$ Also at: Dipartimento DET del Politecnico di Torino, Turin, Italy\\
$^{\rm V}$ Also at: Department of Applied Physics, Aligarh Muslim University, Aligarh, India\\
$^{\rm VI}$ Also at: Institute of Theoretical Physics, University of Wroclaw, Poland\\
$^{\rm VII}$ Also at: An institution covered by a cooperation agreement with CERN\\

\section*{Collaboration Institutes}

$^{1}$ A.I. Alikhanyan National Science Laboratory (Yerevan Physics Institute) Foundation, Yerevan, Armenia\\
$^{2}$ AGH University of Science and Technology, Cracow, Poland\\
$^{3}$ Bogolyubov Institute for Theoretical Physics, National Academy of Sciences of Ukraine, Kiev, Ukraine\\
$^{4}$ Bose Institute, Department of Physics  and Centre for Astroparticle Physics and Space Science (CAPSS), Kolkata, India\\
$^{5}$ California Polytechnic State University, San Luis Obispo, California, United States\\
$^{6}$ Central China Normal University, Wuhan, China\\
$^{7}$ Centro de Aplicaciones Tecnol\'{o}gicas y Desarrollo Nuclear (CEADEN), Havana, Cuba\\
$^{8}$ Centro de Investigaci\'{o}n y de Estudios Avanzados (CINVESTAV), Mexico City and M\'{e}rida, Mexico\\
$^{9}$ Chicago State University, Chicago, Illinois, United States\\
$^{10}$ China Institute of Atomic Energy, Beijing, China\\
$^{11}$ Chungbuk National University, Cheongju, Republic of Korea\\
$^{12}$ Comenius University Bratislava, Faculty of Mathematics, Physics and Informatics, Bratislava, Slovak Republic\\
$^{13}$ COMSATS University Islamabad, Islamabad, Pakistan\\
$^{14}$ Creighton University, Omaha, Nebraska, United States\\
$^{15}$ Department of Physics, Aligarh Muslim University, Aligarh, India\\
$^{16}$ Department of Physics, Pusan National University, Pusan, Republic of Korea\\
$^{17}$ Department of Physics, Sejong University, Seoul, Republic of Korea\\
$^{18}$ Department of Physics, University of California, Berkeley, California, United States\\
$^{19}$ Department of Physics, University of Oslo, Oslo, Norway\\
$^{20}$ Department of Physics and Technology, University of Bergen, Bergen, Norway\\
$^{21}$ Dipartimento di Fisica, Universit\`{a} di Pavia, Pavia, Italy\\
$^{22}$ Dipartimento di Fisica dell'Universit\`{a} and Sezione INFN, Cagliari, Italy\\
$^{23}$ Dipartimento di Fisica dell'Universit\`{a} and Sezione INFN, Trieste, Italy\\
$^{24}$ Dipartimento di Fisica dell'Universit\`{a} and Sezione INFN, Turin, Italy\\
$^{25}$ Dipartimento di Fisica e Astronomia dell'Universit\`{a} and Sezione INFN, Bologna, Italy\\
$^{26}$ Dipartimento di Fisica e Astronomia dell'Universit\`{a} and Sezione INFN, Catania, Italy\\
$^{27}$ Dipartimento di Fisica e Astronomia dell'Universit\`{a} and Sezione INFN, Padova, Italy\\
$^{28}$ Dipartimento di Fisica `E.R.~Caianiello' dell'Universit\`{a} and Gruppo Collegato INFN, Salerno, Italy\\
$^{29}$ Dipartimento DISAT del Politecnico and Sezione INFN, Turin, Italy\\
$^{30}$ Dipartimento di Scienze MIFT, Universit\`{a} di Messina, Messina, Italy\\
$^{31}$ Dipartimento Interateneo di Fisica `M.~Merlin' and Sezione INFN, Bari, Italy\\
$^{32}$ European Organization for Nuclear Research (CERN), Geneva, Switzerland\\
$^{33}$ Faculty of Electrical Engineering, Mechanical Engineering and Naval Architecture, University of Split, Split, Croatia\\
$^{34}$ Faculty of Engineering and Science, Western Norway University of Applied Sciences, Bergen, Norway\\
$^{35}$ Faculty of Nuclear Sciences and Physical Engineering, Czech Technical University in Prague, Prague, Czech Republic\\
$^{36}$ Faculty of Physics, Sofia University, Sofia, Bulgaria\\
$^{37}$ Faculty of Science, P.J.~\v{S}af\'{a}rik University, Ko\v{s}ice, Slovak Republic\\
$^{38}$ Frankfurt Institute for Advanced Studies, Johann Wolfgang Goethe-Universit\"{a}t Frankfurt, Frankfurt, Germany\\
$^{39}$ Fudan University, Shanghai, China\\
$^{40}$ Gangneung-Wonju National University, Gangneung, Republic of Korea\\
$^{41}$ Gauhati University, Department of Physics, Guwahati, India\\
$^{42}$ Helmholtz-Institut f\"{u}r Strahlen- und Kernphysik, Rheinische Friedrich-Wilhelms-Universit\"{a}t Bonn, Bonn, Germany\\
$^{43}$ Helsinki Institute of Physics (HIP), Helsinki, Finland\\
$^{44}$ High Energy Physics Group,  Universidad Aut\'{o}noma de Puebla, Puebla, Mexico\\
$^{45}$ Horia Hulubei National Institute of Physics and Nuclear Engineering, Bucharest, Romania\\
$^{46}$ Indian Institute of Technology Bombay (IIT), Mumbai, India\\
$^{47}$ Indian Institute of Technology Indore, Indore, India\\
$^{48}$ INFN, Laboratori Nazionali di Frascati, Frascati, Italy\\
$^{49}$ INFN, Sezione di Bari, Bari, Italy\\
$^{50}$ INFN, Sezione di Bologna, Bologna, Italy\\
$^{51}$ INFN, Sezione di Cagliari, Cagliari, Italy\\
$^{52}$ INFN, Sezione di Catania, Catania, Italy\\
$^{53}$ INFN, Sezione di Padova, Padova, Italy\\
$^{54}$ INFN, Sezione di Pavia, Pavia, Italy\\
$^{55}$ INFN, Sezione di Torino, Turin, Italy\\
$^{56}$ INFN, Sezione di Trieste, Trieste, Italy\\
$^{57}$ Inha University, Incheon, Republic of Korea\\
$^{58}$ Institute for Gravitational and Subatomic Physics (GRASP), Utrecht University/Nikhef, Utrecht, Netherlands\\
$^{59}$ Institute of Experimental Physics, Slovak Academy of Sciences, Ko\v{s}ice, Slovak Republic\\
$^{60}$ Institute of Physics, Homi Bhabha National Institute, Bhubaneswar, India\\
$^{61}$ Institute of Physics of the Czech Academy of Sciences, Prague, Czech Republic\\
$^{62}$ Institute of Space Science (ISS), Bucharest, Romania\\
$^{63}$ Institut f\"{u}r Kernphysik, Johann Wolfgang Goethe-Universit\"{a}t Frankfurt, Frankfurt, Germany\\
$^{64}$ Instituto de Ciencias Nucleares, Universidad Nacional Aut\'{o}noma de M\'{e}xico, Mexico City, Mexico\\
$^{65}$ Instituto de F\'{i}sica, Universidade Federal do Rio Grande do Sul (UFRGS), Porto Alegre, Brazil\\
$^{66}$ Instituto de F\'{\i}sica, Universidad Nacional Aut\'{o}noma de M\'{e}xico, Mexico City, Mexico\\
$^{67}$ iThemba LABS, National Research Foundation, Somerset West, South Africa\\
$^{68}$ Jeonbuk National University, Jeonju, Republic of Korea\\
$^{69}$ Johann-Wolfgang-Goethe Universit\"{a}t Frankfurt Institut f\"{u}r Informatik, Fachbereich Informatik und Mathematik, Frankfurt, Germany\\
$^{70}$ Korea Institute of Science and Technology Information, Daejeon, Republic of Korea\\
$^{71}$ KTO Karatay University, Konya, Turkey\\
$^{72}$ Laboratoire de Physique des 2 Infinis, Ir\`{e}ne Joliot-Curie, Orsay, France\\
$^{73}$ Laboratoire de Physique Subatomique et de Cosmologie, Universit\'{e} Grenoble-Alpes, CNRS-IN2P3, Grenoble, France\\
$^{74}$ Lawrence Berkeley National Laboratory, Berkeley, California, United States\\
$^{75}$ Lund University Department of Physics, Division of Particle Physics, Lund, Sweden\\
$^{76}$ Nagasaki Institute of Applied Science, Nagasaki, Japan\\
$^{77}$ Nara Women{'}s University (NWU), Nara, Japan\\
$^{78}$ National and Kapodistrian University of Athens, School of Science, Department of Physics , Athens, Greece\\
$^{79}$ National Centre for Nuclear Research, Warsaw, Poland\\
$^{80}$ National Institute of Science Education and Research, Homi Bhabha National Institute, Jatni, India\\
$^{81}$ National Nuclear Research Center, Baku, Azerbaijan\\
$^{82}$ National Research and Innovation Agency - BRIN, Jakarta, Indonesia\\
$^{83}$ Niels Bohr Institute, University of Copenhagen, Copenhagen, Denmark\\
$^{84}$ Nikhef, National institute for subatomic physics, Amsterdam, Netherlands\\
$^{85}$ Nuclear Physics Group, STFC Daresbury Laboratory, Daresbury, United Kingdom\\
$^{86}$ Nuclear Physics Institute of the Czech Academy of Sciences, Husinec-\v{R}e\v{z}, Czech Republic\\
$^{87}$ Oak Ridge National Laboratory, Oak Ridge, Tennessee, United States\\
$^{88}$ Ohio State University, Columbus, Ohio, United States\\
$^{89}$ Physics department, Faculty of science, University of Zagreb, Zagreb, Croatia\\
$^{90}$ Physics Department, Panjab University, Chandigarh, India\\
$^{91}$ Physics Department, University of Jammu, Jammu, India\\
$^{92}$ Physics Program and International Institute for Sustainability with Knotted Chiral Meta Matter (SKCM2), Hiroshima University, Hiroshima, Japan\\
$^{93}$ Physikalisches Institut, Eberhard-Karls-Universit\"{a}t T\"{u}bingen, T\"{u}bingen, Germany\\
$^{94}$ Physikalisches Institut, Ruprecht-Karls-Universit\"{a}t Heidelberg, Heidelberg, Germany\\
$^{95}$ Physik Department, Technische Universit\"{a}t M\"{u}nchen, Munich, Germany\\
$^{96}$ Politecnico di Bari and Sezione INFN, Bari, Italy\\
$^{97}$ Research Division and ExtreMe Matter Institute EMMI, GSI Helmholtzzentrum f\"ur Schwerionenforschung GmbH, Darmstadt, Germany\\
$^{98}$ Saga University, Saga, Japan\\
$^{99}$ Saha Institute of Nuclear Physics, Homi Bhabha National Institute, Kolkata, India\\
$^{100}$ School of Physics and Astronomy, University of Birmingham, Birmingham, United Kingdom\\
$^{101}$ Secci\'{o}n F\'{\i}sica, Departamento de Ciencias, Pontificia Universidad Cat\'{o}lica del Per\'{u}, Lima, Peru\\
$^{102}$ Stefan Meyer Institut f\"{u}r Subatomare Physik (SMI), Vienna, Austria\\
$^{103}$ SUBATECH, IMT Atlantique, Nantes Universit\'{e}, CNRS-IN2P3, Nantes, France\\
$^{104}$ Sungkyunkwan University, Suwon City, Republic of Korea\\
$^{105}$ Suranaree University of Technology, Nakhon Ratchasima, Thailand\\
$^{106}$ Technical University of Ko\v{s}ice, Ko\v{s}ice, Slovak Republic\\
$^{107}$ The Henryk Niewodniczanski Institute of Nuclear Physics, Polish Academy of Sciences, Cracow, Poland\\
$^{108}$ The University of Texas at Austin, Austin, Texas, United States\\
$^{109}$ Universidad Aut\'{o}noma de Sinaloa, Culiac\'{a}n, Mexico\\
$^{110}$ Universidade de S\~{a}o Paulo (USP), S\~{a}o Paulo, Brazil\\
$^{111}$ Universidade Estadual de Campinas (UNICAMP), Campinas, Brazil\\
$^{112}$ Universidade Federal do ABC, Santo Andre, Brazil\\
$^{113}$ University of Cape Town, Cape Town, South Africa\\
$^{114}$ University of Houston, Houston, Texas, United States\\
$^{115}$ University of Jyv\"{a}skyl\"{a}, Jyv\"{a}skyl\"{a}, Finland\\
$^{116}$ University of Kansas, Lawrence, Kansas, United States\\
$^{117}$ University of Liverpool, Liverpool, United Kingdom\\
$^{118}$ University of Science and Technology of China, Hefei, China\\
$^{119}$ University of South-Eastern Norway, Kongsberg, Norway\\
$^{120}$ University of Tennessee, Knoxville, Tennessee, United States\\
$^{121}$ University of the Witwatersrand, Johannesburg, South Africa\\
$^{122}$ University of Tokyo, Tokyo, Japan\\
$^{123}$ University of Tsukuba, Tsukuba, Japan\\
$^{124}$ University Politehnica of Bucharest, Bucharest, Romania\\
$^{125}$ Universit\'{e} Clermont Auvergne, CNRS/IN2P3, LPC, Clermont-Ferrand, France\\
$^{126}$ Universit\'{e} de Lyon, CNRS/IN2P3, Institut de Physique des 2 Infinis de Lyon, Lyon, France\\
$^{127}$ Universit\'{e} de Strasbourg, CNRS, IPHC UMR 7178, F-67000 Strasbourg, France, Strasbourg, France\\
$^{128}$ Universit\'{e} Paris-Saclay Centre d'Etudes de Saclay (CEA), IRFU, D\'{e}partment de Physique Nucl\'{e}aire (DPhN), Saclay, France\\
$^{129}$ Universit\`{a} degli Studi di Foggia, Foggia, Italy\\
$^{130}$ Universit\`{a} del Piemonte Orientale, Vercelli, Italy\\
$^{131}$ Universit\`{a} di Brescia, Brescia, Italy\\
$^{132}$ Variable Energy Cyclotron Centre, Homi Bhabha National Institute, Kolkata, India\\
$^{133}$ Warsaw University of Technology, Warsaw, Poland\\
$^{134}$ Wayne State University, Detroit, Michigan, United States\\
$^{135}$ Westf\"{a}lische Wilhelms-Universit\"{a}t M\"{u}nster, Institut f\"{u}r Kernphysik, M\"{u}nster, Germany\\
$^{136}$ Wigner Research Centre for Physics, Budapest, Hungary\\
$^{137}$ Yale University, New Haven, Connecticut, United States\\
$^{138}$ Yonsei University, Seoul, Republic of Korea\\
$^{139}$  Zentrum  f\"{u}r Technologie und Transfer (ZTT), Worms, Germany\\
$^{140}$ Affiliated with an institute covered by a cooperation agreement with CERN\\
$^{141}$ Affiliated with an international laboratory covered by a cooperation agreement with CERN.\\

\end{flushleft} 
  
\end{document}